\documentclass[10pt,preprint]{aastex}
\usepackage{natbib,natbibspacing}

\slugcomment{To appear in the Astrophysical Journal.}
\shorttitle{Predicting PMS star magnetic fields}
\shortauthors{Gregory et al.}

\begin{document}

\title{Can we predict the global magnetic topology of a pre-main sequence star \\from its position in the Hertzsprung-Russell diagram?}

\author{S.~G.~Gregory\altaffilmark{1}, J.-F.~Donati\altaffilmark{2}, J.~Morin\altaffilmark{3,4}, G.~A.~J.~Hussain\altaffilmark{5}\\
N.~J.~Mayne\altaffilmark{6}, L.~A.~Hillenbrand\altaffilmark{1}, M.~Jardine\altaffilmark{7}}

\altaffiltext{1}{California Institute of Technology, MC 249-17, Pasadena, CA 91125}
\email{sgregory@caltech.edu}
\altaffiltext{2}{IRAP-UMR 5277, CNRS $\&$ Universit{\'e} de Toulouse, 14 Av. E. Belin, F-31400 Toulouse, France}
\altaffiltext{3}{Universit{\"a}t G{\"o}ttingen, Institut f{\"u}r Astrophysik, Friedrich-Hund-Platz 1, D-37077 G{\"o}ttingen, Germany}
\altaffiltext{4}{Dublin Institute for Advanced Studies, School of Cosmic Physics, 31 Fitzwilliam Place, Dublin 2, Ireland}
\altaffiltext{5}{ESO, Karl-Schwarzschild-Str. 2, D-85748 Garching, Germany}
\altaffiltext{6}{School of Physics, University of Exeter, Exeter, EX4 4QL, United Kingdom}
\altaffiltext{7}{School of Physics and Astronomy, University of St Andrews, St Andrews, KY16 9SS, United Kingdom}

\begin{abstract}
Zeeman-Doppler imaging studies have shown that the magnetic fields of T Tauri stars can be significantly 
more complex than a simple dipole and can vary markedly between sources.  We collect and summarize the magnetic field 
topology information obtained to date and present Hertzsprung-Russell (HR) diagrams for the stars in the sample.  Intriguingly, the large 
scale field topology of a given pre-main sequence (PMS) star is strongly dependent upon the stellar internal structure, with the strength of the 
dipole component of its multipolar magnetic field decaying rapidly with the development of a radiative core.
Using the observational data as a basis, we argue that the general characteristics of the global magnetic field of a 
PMS star can be determined from its position in the HR diagram.  Moving from hotter and more luminous to cooler and less 
luminous stars across the PMS of the HR diagram, we present evidence for four distinct magnetic 
topology regimes.  Stars with large radiative cores, empirically estimated to be those with a core mass in excess
of $\sim$40\% of the stellar mass, host highly complex and dominantly non-axisymmetric magnetic fields, while those with smaller radiative
cores host axisymmetric fields with field modes of higher order than the dipole dominant (typically, but not always, the octupole).  Fully
convective stars stars above $\gtrsim0.5\,{\rm M}_\odot$ appear to host dominantly axisymmetric 
fields with strong (kilo-Gauss) dipole components.  
Based on similarities between the magnetic properties of PMS stars and main sequence M-dwarfs with similar internal structures, 
we speculate that a bistable dynamo process operates for lower mass stars ($\lesssim0.5\,{\rm M}_\odot$ at an age of a few Myr) and 
that they will be found to host a variety of magnetic field topologies.  If the magnetic topology trends across the HR
diagram are confirmed they may provide a new method of constraining PMS stellar evolution models.
\end{abstract}



\section{Introduction}\label{intro}
At the end of the protostellar phase of spherical accretion, a newly formed and optically visible pre-main sequence (PMS) T Tauri 
star is highly luminous due to its large surface area ($L_\ast\propto R_\ast^2$).  The contracting star thus begins its journey 
towards the main sequence in the upper right of the Hertzsprung-Russell (HR) diagram while accreting material from its 
circumstellar disk.  At this stage the temperature $T$ and density $\rho$ in the central regions of the star are not sufficient for 
thermonuclear reactions to occur, and the stellar luminosity is supplied by the release of the gravitational potential energy via the 
stellar contraction.  During the fully convective phase of evolution the PMS star follows an almost vertical downward path in the 
HR diagram, called the Hayashi track \citep{hay61}.  

As the gravitational contraction proceeds the opacity $\kappa$ in the central regions becomes dominated 
by free-free and bound-free transitions (e.g. \citealt{war11}), for which $\kappa\propto\rho T^{-7/2}$.  As the 
temperature continues to rise the central opacity thus drops, the star becomes more transparent, and
the radiative gradient decreases below the critical value required to support convection (see the discussion in \citealt{harbook})
and a radiative core forms.  This radiative core continues to grow reducing the 
depth of the convective zone.\footnote{More massive T Tauri stars become entirely radiative during their PMS 
evolution and some develop convective cores on the main sequence if the power generated from thermonuclear reactions is 
sufficient (see e.g. the models of \citealt{lej01}).  Stars with mass $\lesssim$0.35$\,{\rm M}_\odot$ arrive on the main sequence 
and hydrogen fusion begins before a radiative core can develop, and thus retain a fully convective interior during their PMS 
evolution \citep{cha97}.}  Eventually the temperature and luminosity of the contracting star rise, and it leaves its Hayashi track and moves 
onto its Henyey track \citep{hen65}, a process sometimes referred to as the 
``convective-radiative transition'' (e.g. \citealt{may10}).  The rapid increase in 
effective temperature at a slowly increasing luminosity during the Henyey phase leads to a clear ``gap'' 
in color-magnitude diagrams of PMS clusters \citep{may07}.  The size of the gap is dependent on the mass cut-off between fully
and partially convective stars which itself is a function of stellar age, meaning that the gap can, in principle, be used 
as a distance independent age indicator \citep{may08}.    

Several observational results have been attributed to the development of a 
radiative core at the end of the fully convective phase of evolution.  \citet{reb06} argue that the ratio
of X-ray to bolometric luminosity is systematically lower for stars with radiative cores compared to those  
on the fully convective portion of their mass tracks, with \citet{may10} finding similar reductions in 
X-ray luminosities in older PMS clusters (the older the cluster the greater the fraction of stars that have ended
the fully convective phase).  \citet{ale12} invoke radiative core development to
explain the reduction in the scatter in X-ray luminosities apparent in rotation-activity plots in older PMS star forming regions.   
Furthermore, the growth of a radiative core appears to coincide with a reduction 
in the number of periodically variable T Tauri stars \citep{sau09}.  The authors attribute this result to differing
cool spot distributions in fully and partially convective stars with large cool spots (where bundles of magnetic flux burst through the 
stellar surface into the atmosphere) on fully convective objects and 
smaller more numerous spots on stars with radiative cores which naturally leads to less rotationally modulated
variability.  All of these observational results can be qualitatively explained if the external magnetic field topology of T Tauri stars
changes as the stellar internal structure transitions from fully to partially convective.  

T Tauri stars have long been known to possess surface-averaged magnetic fields of order a kilo-Gauss as determined from
Zeeman broadening measurements (e.g. \citealt{joh99a,joh07,yan11} and references therein).  
Such strong magnetic fields can disrupt circumstellar disks at a distance of a few stellar radii \citep{kon91}, provided that they are 
sufficiently globally ordered, a key requirement of magnetospheric 
accretion models (see \citealt{gre10} for a review).  The disk truncation radius $R_t$ is set by the interplay between the 
strength of the stellar magnetosphere at the inner disk, which can be approximated from the polar strength of the dipole 
component of the multipolar stellar magnetic field at the surface of the star $B_{\rm dip}$, and the disk mass accretion rate 
$\dot{M}$.  Larger disk truncation radii are expected for stronger dipole components and/or weaker mass accretion rates 
as $R_t \propto B_{\rm dip}^{4/7}\dot{M}^{-2/7}$ (e.g. \citealt{kon91}), quantities which may vary significantly with time.  
Typical disk truncation radii are believed to be $\sim$5$\,R_\ast$ \citep{gul98}, or $\sim$0.05$\,{\rm AU}$ 
for a prototypical $2\,{\rm R}_{\odot}$ T Tauri star.  Such small scales, within or comparable to typical dust sublimation radii, can be 
probed with high resolution spectroscopy of gas emission lines (e.g. \citealt{naj03}) and by long baseline 
interferometry (see \citealt{mil07}, \citealt{ake08} and \citealt{mil07} for reviews of the technique).  Gas is typically found to 
extend closer towards the star than the dusty component of the disk for both T Tauri stars and the related more massive 
Herbig Ae/Be stars \citep{kra08,ise08,eis09,eis10,rag09}.  For some sources the gas component of the disk provides a 
significant amount of the detected inner disk flux \citep{ake05}.  The inner disk gas couples to the field
lines of the stellar magnetosphere and is channeled onto the stellar surface at high velocity, where it shocks and produces
detectable hotspots that are the source of continuum emission in excess of the stellar photospheric emission, as well 
as soft X-rays (e.g. \citealt{cal98,kas02,arg11a,arg11b}).  The geometry and distribution of accretion hot spots is a strong 
function of the stellar magnetospheric geometry \citep{rom04a,gre05,gre06a,moh08}.

Complementing the Zeeman broadening analysis, which is carried out 
in unpolarized light, measurement of the level of circular polarization in both accretion related emission lines and 
photospheric absorption lines allows information to be derived about the field topology
itself \citep{val04,don07,don08b,don10b}.  The large scale magnetic fields of accreting T Tauri stars appear to be well-ordered, and 
are simpler than the complex and loopy surface field regions (e.g. \citealt{val04}).  However, although the large-scale field that is interacting
with the inner disk is somewhat dipole-like in appearance, the path of field lines close to the star, and consequently the magnetospheric 
accretion flow, is distorted close to the stellar surface by the complex field regions \citep{gre08,ada11}.

Magnetic surface maps have now been published for a number of accreting T Tauri stars 
\citep{don07,don08b,don10a,don10b,don11a,don11c,don11b,don12,hus09,ske11}, one non-accreting weak line T Tauri star \citep{ske10},
and a few older post T Tauri stars \citep{dun08,mar11,wai11}, derived from the technique of Zeeman-Doppler imaging, as we discuss in 
the following section.  Most of the published magnetic maps have been obtained as part of the Magnetic Protostars and Planets (MaPP) project.  
The main goal of this large program with the ESPaDOnS spectropolarimeter at the Canada-France-Hawai'i telescope \citep{don03}, 
and the twin instrument NARVAL at the T{\'e}lescope Bernard Lyot in the Pyren{\'e}es \citep{aur03}, is to investigate variations in the 
magnetic topology of accreting T Tauri stars of different mass\footnote{In this paper we refer to stars of mass 
$M_\ast\lesssim0.5\,{\rm M}_\odot$ as low mass, $0.5\lesssim M_\ast/{\rm M}_\odot\lesssim1.0$ as intermediate mass, 
and $M_\ast\gtrsim1.0\,{\rm M}_\odot$ as high mass PMS stars.}, 
age, accretion rate, rotation period, and outflow properties (see \citealt{don10b} for a brief introduction to the program).  

The initial MaPP results have demonstrated that accreting T Tauri stars possess multipolar magnetic fields but with dipole components
that are strong enough to disrupt the inner disk at distances of up to several stellar radii.  Intriguingly, the field complexity and 
the polar strength of the dipole component appear to increase and decrease respectively when comparing stars with fully convective interiors
to those which have developed radiative cores (e.g. \citealt{don11c}) - a concept that we explore fully in this paper.  Similar variations in magnetic field topology
of main sequence (MS) M-dwarfs that span the fully convective divide have been discovered by \citet{mor08,mor10} and \citet{don08a}.     

\begin{figure}[!t]
   \centering
    \includegraphics[width=65mm]{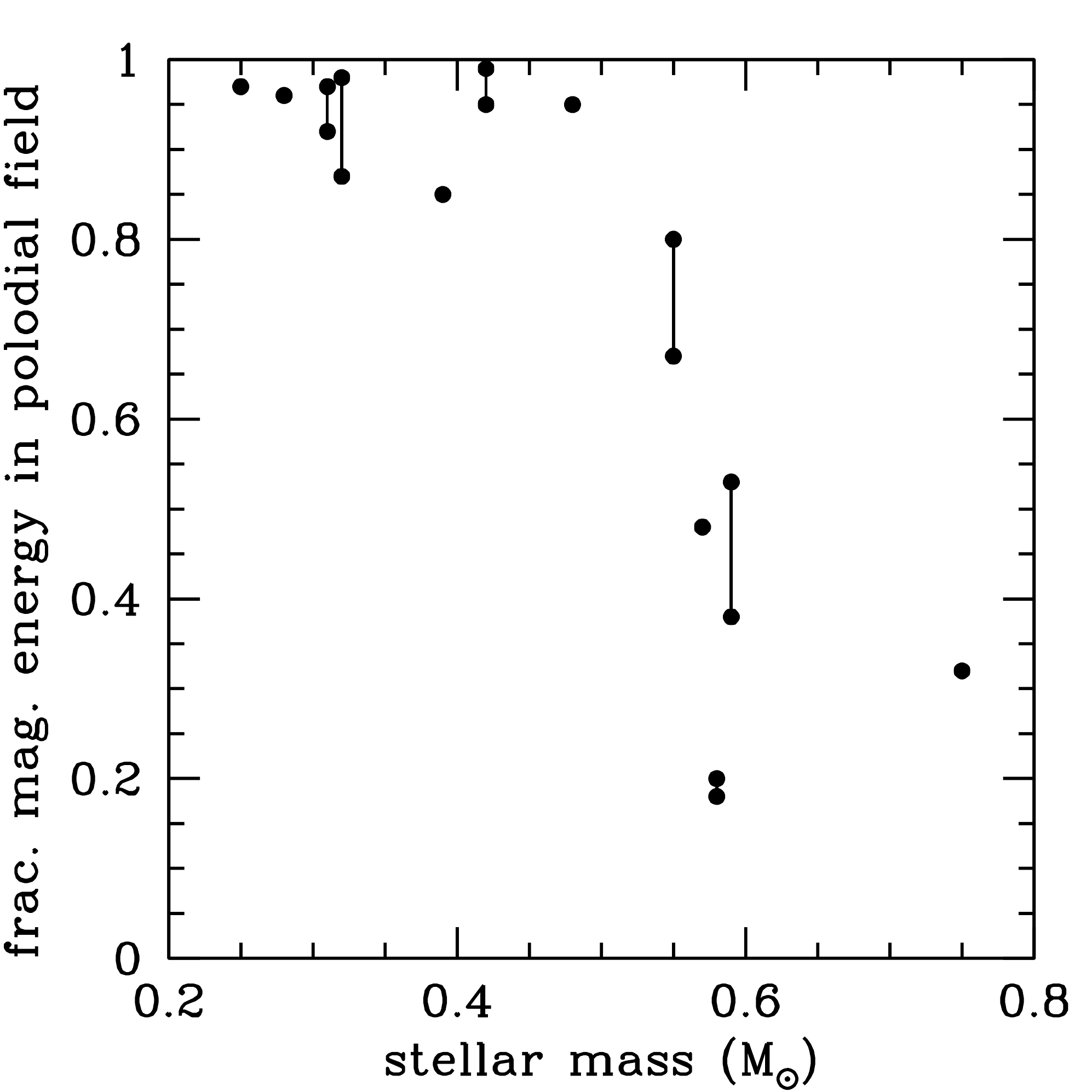}
   \caption{The magnetic energy in the poloidal field modes relative to the magnetic energy in all field modes, as a function 
                of the mass of main sequence M-dwarfs, determined via a spherical harmonic decomposition of the magnetic maps derived    
                from ZDI studies (data presented in \citealt{mor08} and \citealt{don08a}).  Points joined by vertical bars represent 
                stars observed at two different epochs.  For 
                stars with small radiative cores (those above $\sim$0.35$\,{\rm M}_{\odot}$; \citealt{cha97}) the field remains dominantly poloidal
                until $\sim$0.5$\,{\rm M}_\odot$.  For more massive stars with large radiative cores strong toroidal field components develop 
                with increasing stellar mass and the fractional magnetic energy in the poloidal field decreases.  This increase in field 
                complexity with the size of a radiative core results in the dipole component of the field becoming less significant relative
                to the higher order field components.  Completely convective stars have simpler, mostly poloidal, 
                large scale fields, although very low mass 
                fully convective M-dwarfs (below $\sim$0.2$\,{\rm M}_\odot$) are also capable of hosting complex magnetic fields (see \citealt{mor10}; 
                data points not shown here).}
   \label{mdwarfs}
\end{figure}

\begin{figure}[!t]
  \centering
    \includegraphics[width=65mm]{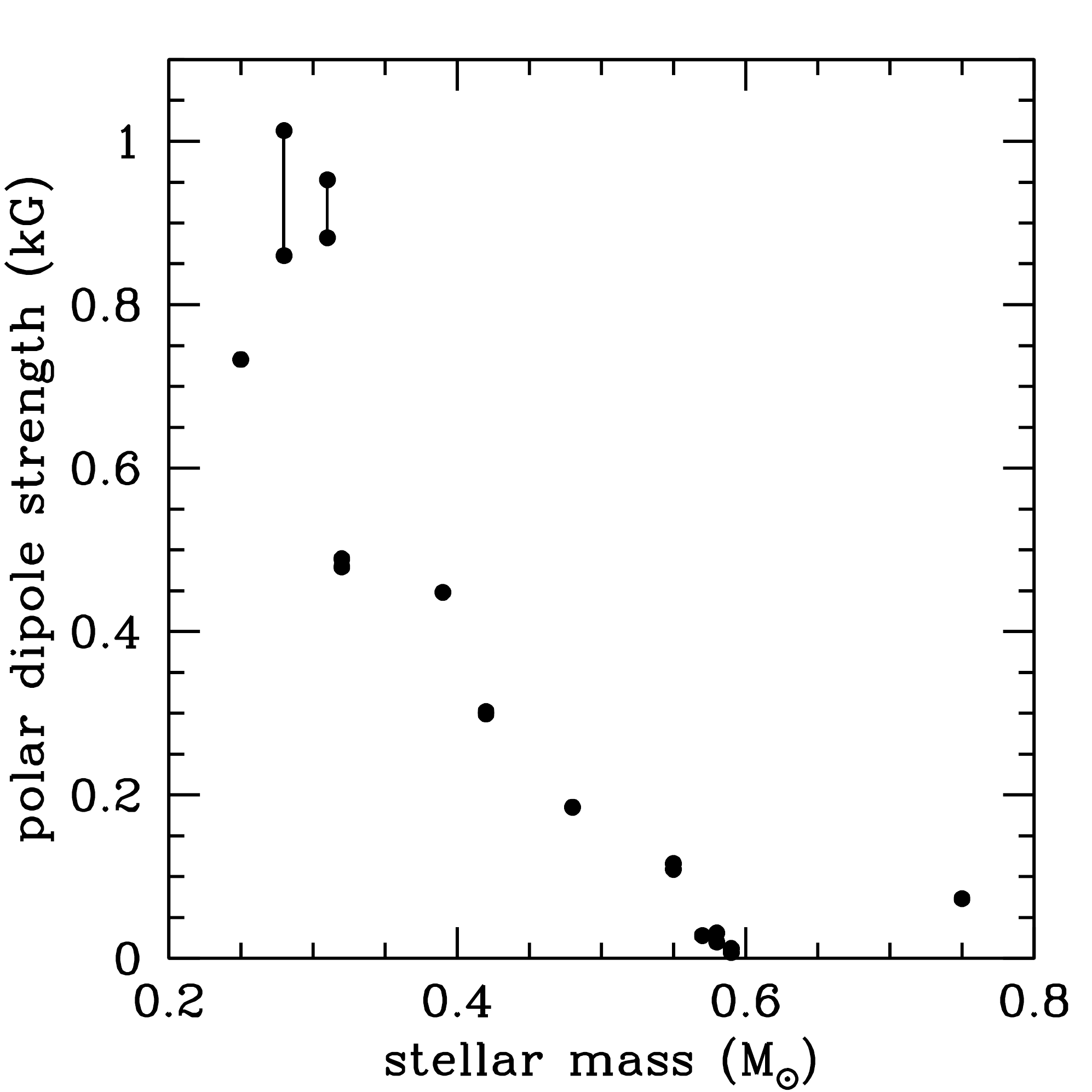}
  \caption{The decay of the polar strength of the $m=0,\ell=1$ dipole component $B_{\rm dip}$ for the main sequence M-dwarf stars observed by 
                \citet{mor08} and \citet{don08a}.  Points connected by vertical bars represent stars observed at two different epochs.    
               Low mass completely convective stars have strong dipole components.  As the stellar mass, and therefore the size 
               (volume) and mass of the radiative core, increases above the limit for full convection the strength of the dipole component 
               decays rapidly.  
               }
  \label{bdip_mdwarf}
\end{figure}

On the MS stars below $\sim$0.35$\,{\rm M}_\odot$, later than roughly spectral type M4, have a fully convective 
internal structure while more massive stars do not \citep{cha97}.  Fully convective M-dwarfs close to the fully convective limit 
($\gtrsim0.2\,{\rm M}_\odot$) host simple, axisymmetric magnetic fields with strong dipole components \citep{mor08}.  M-dwarfs of earlier 
spectral type which are partially convective with small radiative cores have magnetic fields with weaker dipole components that are dominantly
axisymmetric (most of the magnetic energy is in the poloidal field modes; \citealt{don08a}).  M-dwarfs with more substantial radiative 
cores have both weak dipole components and more complex magnetic fields (less magnetic energy in the poloidal field modes; \citealt{don08a}).  
This behavior is illustrated in Figures \ref{mdwarfs} and \ref{bdip_mdwarf}.
Clear differences in (the large scale) stellar magnetic field topologies are observed 
as a direct manifestation of the differing stellar internal structure and therefore, presumably, the different type of dynamo mechanism 
operating in fully and partially convective stars.  Stars with outer convective zones and radiative cores are believed to possess a solar-like 
tacholine, a shear layer between the core and envelope, whereas fully convective stars lack this interface, and the dynamo process is 
different (e.g. \citealt{bro08}). 
  
\citet{mor11}, following on from \citet{gou08}, point out that the simple large scale 
fields of many fully convective M-dwarfs are more akin to the simple magnetic topologies of the gas giant planets within our Solar System 
(e.g. \citealt{wil82}), rather than the messy and complex fields observed on active zero-age main sequence K-type stars (e.g. \citealt{don03zams}).  
However, the lowest mass M-dwarfs ($\lesssim$0.2$\,{\rm M}_\odot$) with very similar stellar parameters (rotation rate and mass)
can show drastically different magnetic field topologies \citep{mor10}.  This may be caused by a 
bistable dynamo process, with a weak and a strong field dynamo branch, such that the two different dynamo regimes co-exist over 
a certain range of parameters (see figure 4 of \citealt{mor11}, adapted from \citealt{rob88}).  Fully convective stars in the strong
field regime maintain steady and simple axisymmetric dominantly dipolar magnetic fields while fully convective stars with identical 
stellar parameters but in the weak field regime host complex multipolar fields which evolve rapidly in time and have weak dipole 
components.  

In this paper we explore the trends in the magnetic field topology of PMS stars with stellar internal structure that are now emerging
from spectropolarimetric observing programs.  Unlike MS M-dwarfs, however, the internal structure of PMS stars is changing rapidly 
due to the stellar contraction and (if the star is massive enough)
the development of a radiative core.  The stellar age is thus an important parameter in addition to mass when examining 
magnetic field topology variations for PMS stars caused by changes in the internal structure of the star. 

Given the importance of the magnetic field 
topology in controlling the star-disk interaction, in \S\ref{topo} we summarize
the magnetic topology information, and the fundamental stellar parameters, that have been obtained to date for accreting PMS stars.  
In \S\ref{hrdiagrams} we position the stars with derived magnetic maps onto HR diagrams constructed from two different PMS 
evolutionary models, examine the role of the development of a radiative core in 
setting the magnetic field topology, and compare the field topologies of PMS stars to those of MS M-dwarfs with similar 
internal structures.  In \S\ref{predictHRD} we argue that it may be possible to predict the global magnetic topology characteristics
of a given star (e.g. whether or not the field will have a strong dipole component, is dominantly axisymmetric or non-axisymmetric 
etc) based on its position in the HR diagram.  Our ability to do this, however, is dependent on a number of caveats and assumptions
including the accuracy with which the star can be positioned in the HR diagram observationally, and the veracity of the PMS evolutionary 
models themselves.  In \S\ref{discussion} we explore the implications of magnetic topology variations in terms of the star-disk interaction, 
while \S\ref{conclusions} contains our conclusions.


\section{T Tauri magnetic field topology}\label{topo}
Over the past few years spectropolarimetric Zeeman-Doppler imaging (ZDI) studies have revealed that the field topology of 
T Tauri stars can vary significantly between sources (\citealt{don01} and \citealt{don09} provide reviews of the basic methodology of 
ZDI, while details specific to accreting T Tauri stars are discussed in \citealt{don10b} and \citealt{hus12}). 
Magnetic maps of T Tauri stars are constructed by measuring the circular polarization (Stokes V) signal in both accretion related 
emission lines and in photospheric absorption lines, over at least one complete stellar rotation cycle and in practice several cycles.  
Circular polarization can be measured directly in the accretion related emission lines, for example HeI 
5876{\AA } \citep{joh99b,don08b} or the CaII infrared triplet \citep{don07}, but the signal is often too weak in a given individual 
magnetically sensitive photospheric absorption line.  Thus, cross-correlation techniques, such as least-squares 
deconvolution, are used to extract information from as many lines as possible \citep{don97}.  
By monitoring how rotationally modulated distortions, generated by magnetic surface features, move through the Stokes V profile,
a 2D distribution of magnetic polarities across the surface of stars can be determined 
using maximum entropy reconstruction techniques \citep{bro91}, as well as the field orientation within the magnetic 
regions \citep{db97}.

The ability to derive maps of the surface magnetic topology of T Tauri stars (and of stars generally) is subject to 
some limitations, as discussed in detail by \citet{hus12}.  ZDI, like all polarization techniques, suffers from the effects of flux 
cancellation.  Photons received from regions of the stellar atmosphere permeated by opposite polarity magnetic fields are 
polarized in the opposite sense.  Their signals
can therefore cancel, resulting in a net polarization signal of zero.  Due to this flux cancellation effect it is possible to recover 
information only about the medium-to-large scale field topology.  Detailed features on scales that can be resolved 
in solar magnetograms remain below the resolution limit achievable in stellar magnetic maps (see \citealt{gre10} for further discussion).  
Spectropolarimetric Stokes V studies thus likely miss a large fraction of the total magnetic flux \citep{rei09}, presumably contained
within the tangled and complex small scale field, perhaps on the scale of bipolar groups detected on the Sun.  The resolution 
achievable in stellar magnetic maps is also dependent on the stellar rotation period and the inclination.      

Once magnetic maps have been derived the field topology can be reconstructed as the values of the 
coefficients of a spherical harmonic decomposition, and the strength of the various field modes determined (e.g. \citealt{don06}).  For example, it
is possible to decompose the field into the sum of a poloidal plus a toroidal component, and to calculate the strength, 
tilt, and phase of tilt, of the various multipole moments \citep{don10b}.  Due to the stellar inclination surface magnetic field information 
cannot be obtained across portions of the stellar surface that remain hidden from view to an observer.  This limitation is important
when constructing 3D models of T Tauri magnetospheres via field extrapolation from the magnetic maps 
(e.g. \citealt{gre08}) as an assumption must be made whether to favor the symmetric (the odd $\ell$ number modes e.g. the dipole, the octupole, 
the dotriacontapole etc) or antisymmetric (the even $\ell$ number modes e.g. the quadrupole, the hexadecapole etc) field modes.  As 
part of the tomographic imaging process maps of the surface distribution of cool (dark) spots and accretion related
hot spots are also derived \citep{don10b}.  These maps suggest that for the bulk of accreting T Tauri stars gas in accretion 
columns impacts the stellar surface at high latitudes close to the poles.  This suggests that it is antisymmetric field modes, like the dipole and 
the octupole, that dominate.  If the symmetric field modes were to dominate then the majority of the gas would accrete onto the equatorial 
regions, for example \citet{lon07}, which is not observed.  The choice of whether to favor the symmetric or anti-symmetric modes, although
important when constructing 3D models of the stellar magnetosphere, does not fundamentally change the appearance of the 
magnetic field maps \citep{hus12}.

Magnetic maps have now been published for a number of T Tauri stars (see \citealt{gredon11} for a review).  Some T Tauri stars 
host simple axisymmetric large-scale magnetic fields that are dominantly dipolar (AA~Tau and BP~Tau) or where a higher order field 
mode dominates (typically, but not always, the octupole; V2129~Oph, GQ~Lup, TW~Hya, and MT~Ori), while others host highly 
complex magnetic fields that are dominantly non-axisymmetric 
with many high order multipole components (V4046~Sgr~AB, CR~Cha, CV~Cha, and V2247~Oph).  In 
Appendix \ref{starinfo} we provide detailed information about the magnetic field topology of every accreting T Tauri star for which
magnetic maps have been derived to date, with the main stellar properties listed in Tables \ref{params_fund} and \ref{params}. The 
stellar internal structure information, masses, and ages have been derived by placing the stars onto the HR diagram as discussed 
in \S\ref{new3.1}.

\begin{deluxetable}{lcccccccc}
\tabletypesize{\scriptsize}
\tablewidth{0pt}
\tablecaption{Fundamental parameters of accreting T Tauri stars with observationally derived magnetic maps.\label{params_fund}}
\tablehead{
\colhead{Star}                                                       		&
\colhead{Spec. Type}                                            		& 
\colhead{$T_{\rm eff}$ (K)\tablenotemark{b}}                      & 
\colhead{$\log(L_\ast$/L$_\odot)$\tablenotemark{b}} 	& 
\colhead{$P_{\rm rot}$ (d)}						&
\colhead{$\tau_c$ (d)}							&
\colhead{Binary?}	 							&
\colhead{Separation (AU)}						 	&
\colhead{refs}}                                                                     
\startdata
\cutinhead{Stars with strong dipole components and dominantly axisymmetric large scale magnetic fields}
AA Tau 					& K7 & 4000$\pm$100	& 0.0$\pm$0.1 		& 8.22 & 230  & no & $\ldots$ & 1\\
BP Tau 					& K7 & 4055$\pm$112  	& -0.03$\pm$0.1 	& 7.6   & 237 & no & $\ldots$ & 2\\
\cutinhead{Stars with dominant high order magnetic field components ($\ell>1$) and axisymmetric large scale magnetic fields}
V2129 Oph\tablenotemark{c} 	& K5 & 4500$\pm$100    	& 0.15$\pm$0.1 	& 6.53 & 182 & yes & 78 & 3,4\\
GQ Lup					& K7 & 4300$\pm$50       &  0.0$\pm$0.1          & 8.4   & 199 & yes & 100 & 5 \\
TW Hya 					& K7 & 4075$\pm$75    	& -0.46$\pm$0.1 	& 3.56 & 180 & no & $\ldots$ & 6\\
MT Ori  					& K2 & 4600$\pm$100   	& 1.49$\pm$0.13 	& 8.53 & 322 & no & $\ldots$ & 7\\
\cutinhead{Stars with complex non-axisymmetric large scale magnetic fields and weak dipole components}
V4046 Sgr A   				& K5 & 4370$\pm$100	& -0.39$\pm$0.1 	& 2.42 & 117 & yes & 0.041 & 8\\
V4046 Sgr B   				& K5 & 4100$\pm$100	& -0.57$\pm$0.1 	& 2.42 & 130 & yes & 0.041 & 8\\
CR Cha   					& K2 & 4900$\pm$100 	& 0.58$\pm$0.13 	& 2.3   & 92 & no & $\ldots$ & 9\\
CV Cha  					& G8 & 5500$\pm$100   	& 0.89$\pm$0.08 	& 4.4   & 56 & yes & 1596 & 9\\
V2247 Oph\tablenotemark{d} 	& M1 & 3500$\pm$150    	& -0.33$\pm$0.1 	& 3.5   & 222 & yes & 36 & 10\\ 
\enddata
\tablenotetext{a}{Cols. (1-2): star name and observation date.  Cols. (3-6): effective temperature, luminosity, rotation period, and an 
                       estimate of the local convective turnover time discussed in \S\ref{tau_c}.  Cols. (7-8): binary star status and separation.  Col. (9): reference where
                         effective temperature and luminosity assignment is discussed.}
\tablenotetext{b}{Error estimates are discussed in \S\ref{topo}, fifth and sixth paragraphs.}
\tablenotetext{c}{The V2129~Oph luminosity was updated from that used in \citet{don07} by \citet{don11a} using a more refined
                           distance estimate to the $\rho$-Oph star forming region of $120\,{\rm pc}$ \citep{loi08}.}
\tablenotetext{d}{The luminosity of V2247~Oph has been updated from \citet{don10a} using the \citet{loi08} distance estimate.}                           
\tablerefs{
(1) \citealt{don10b}; (2) \citealt{don08b}; (3) \citealt{don07}; (4) \citealt{don11a}; (5) \citealt{don12}; (6) \citealt{don11c}; 
(7) \citealt{ske11};  
(8) \citealt{don11b}; (9) \citealt{hus09}; (10) \citealt{don10a}. }
\end{deluxetable}


\begin{deluxetable}{llccccccccc}
\tabletypesize{\scriptsize}
\rotate
\tablewidth{0pt}
\tablecaption{Parameters of accreting T Tauri stars with observationally derived magnetic maps.\label{params}}
\tablehead{
&&&\citet{sie00}&&&\citet{tog11}&&&& \\
\colhead{Star}                                               	&
\colhead{Date}                                              	& 
\colhead{$M_\ast$/M$_\odot$}                     	& 
\colhead{$M_{\rm core}$/M$_\ast$}             	&
\colhead{age (Myr)}                                      	& 
\colhead{$M_\ast$/M$_\odot$}                     	& 
\colhead{$M_{\rm core}$/M$_\ast$}             	&
\colhead{age (Myr)}                                     	&
\colhead{$B_{\rm dip}$ (kG)\tablenotemark{b}}	&  
\colhead{refs} 
}
\startdata
\cutinhead{Stars with strong dipole components and dominantly axisymmetric large scale magnetic fields}
AA Tau 					& Dec07		& 0.70 & 0.00  & 1.42    	 & 0.63 & 0.00 & 1.39 &  1.9 & 1\\
$\ldots$ 					& Jan09		& $\ldots$ & $\ldots$ & $\ldots$ & $\ldots$ & $\ldots$ & $\ldots$ & 1.7 & 1 \\
BP Tau\tablenotemark{c} 		& Feb06		& 0.75 & 0.00  & 1.80    	 & 0.69 & 0.00 & 1.64 & 1.2 & 2\\
$\ldots$					& Dec06		& $\ldots$ & $\ldots$ & $\ldots$ & $\ldots$ & $\ldots$ & $\ldots$ & 1.0 & 2 \\
\cutinhead{Stars with dominant high order magnetic field components ($\ell>1$) and axisymmetric large scale magnetic fields}
V2129 Oph\tablenotemark{d} 	& Jun05 		& 1.36 & 0.19 & 3.67  	 & 1.14 & 0.10 & 2.28 &  0.3 & 3,4\\
$\ldots$	  				& Jul09 		& $\ldots$ & $\ldots$ & $\ldots$        	 & $\ldots$ & $\ldots$ & $\ldots$ &  1.0 & 4\\
GQ Lup					& Jul09		& 1.06 & 0.13  & 3.33		 & 0.93   & 0.02   & 2.39   &  1.1 & 5\\
$\ldots$					& Jun11		& $\ldots$  & $\ldots$  & $\ldots$		 & $\ldots$   & $\ldots$    & $\ldots$   &  0.9 & 5\\
TW Hya 					& Mar08 		& 0.83 & 0.18 & 9.17       	 & 0.84 & 0.27 & 7.13 &  0.4 & 6\\
$\ldots$ 					& Mar10 		& $\ldots$ & $\ldots$ & $\ldots$ 	 & $\ldots$ & $\ldots$ & $\ldots$ &  0.7 & 6\\
MT Ori  					& Dec08 		& 2.7   & $>0.03$,$<0.36$ & 0.24 & 1.96 & 0.00 & 0.18 & $<$0.1 & 7\\
\cutinhead{Stars with complex non-axisymmetric large scale magnetic fields and weak dipole components}
V4046 Sgr A  				& Sep09 		& 0.91 & 0.47 & 13.0  	 & 0.98 & 0.64 & 12.0 &  0.1 &  8\\
V4046 Sgr B  				& Sep09 		& 0.87 & 0.40 & 13.0  	 & 0.85 & 0.50 & 12.1 &  0.08 &  8\\
CR Cha  					& Apr06  		& 1.96 & 0.65 & 2.89       	 & 1.78 & 0.39 & 1.67 &  $>$0.09 &  9\\
CV Cha  					& Apr06  		& 2.04 & 0.98 & 4.51       	 & 2.19 & 0.94 & 2.97 &  $>$0.02 &  9\\
V2247 Oph\tablenotemark{e} 	& Jul08 		& 0.36 & 0.00 &  1.4  	 & 0.35 & 0.00 & 1.67 &  0.1 &  10\\ 
\enddata
\tablenotetext{a}{Cols. (3-5): stellar mass, radiative core mass relative to the stellar mass and 
                        the age derived from the models of \citet{sie00}.  Cols. (6-8): as cols. (3-5) but from the models of \citet{tog11}.  Col. (9): 
                        rotation period. 
                       Cols. (10-11): the polar strength of the dipole $B_{\rm dip}$ field component with
                       reference.}
\tablenotetext{b}{All of the stars host multipolar magnetic fields, but we list only the dipole component given its importance to the star-disk interaction.  
                The large-scale magnetic fields of AA~Tau, BP~Tau, V2129~Oph and TW~Hya are well described by a tilted dipole plus a tilted 
                octupole field component \citep{gredon11}, as is the field of GQ~Lup \citep{don12}.}                       
\tablenotetext{c}{The magnetic parameters for BP~Tau are based on an old version of the magnetic imaging code 
                          and will be updated in a forthcoming paper, see \S\ref{bptau}.}
\tablenotetext{d}{The Jun05 V2129~Oph data presented in \citet{don07} was reanalyzed by \citet{don11a}.}
\tablenotetext{e}{Parameters for V2247~Oph updated from \citet{don10a}.}   
\tablerefs{
(1) \citealt{don10b}; (2) \citealt{don08b}; (3) \citealt{don07}; (4) \citealt{don11a}; (5) \citealt{don12}; (6) \citealt{don11c}; 
(7) \citealt{ske11}; (8) \citealt{don11b}; (9) \citealt{hus09}; (10) \citealt{don10a}. }
\end{deluxetable}

The effective temperatures $T_{\rm eff}$ and luminosities $L_\ast$ listed in 
Table \ref{params_fund} are the values that were adopted in each of the papers where the magnetic maps were published, as
listed in the reference column of the table, and to which readers are referred for detailed discussion.  The one exception is the 
luminosity of V2247~Oph which we have updated using a more refined estimate of the distance to the $\rho$-Oph star forming 
region (see \S\ref{v2247}).  Typically the effective temperatures were sourced from previously published literature 
values as derived from high resolution spectra.  The exception to this is GQ~Lup for which \citet{don12} derived a new 
$T_{\rm eff}$ as previous literature values were estimated from low resolution spectra and proved to be highly discrepant.  
As a consistency check a new spectral classification tool (called MagIcS) has been developed
and applied to the ESPaDOnS spectra and has been tested for a number of main sequence and PMS template stars (Donati, in prep.). 
The assumed errors in $T_{\rm eff}$ values are also taken from the previously published values, or assumed to be 
$100\,{\rm K}$ when no error estimate is available (errors from the spectral classification tool are $<100\,{\rm K}$).    

Luminosities were derived from the visual magnitudes and distance estimates to the various star forming regions, taking account of the 
uncertainty associated with the presence of surface cool spots.  In the papers with published magnetic maps the error in $\log(L_\ast/{\rm L}_\odot)$ is typically 
assumed to be $0.1\,{\rm dex}$, with the exception of CR~Cha, CV~Cha and MT~Ori.  Further stellar parameters, including the stellar 
rotation periods adopted during the magnetic map reconstruction process, are listed in Table \ref{params}.  Given the importance of the 
dipole component in controlling the star-disk interaction, see \S\ref{dipcomp}, we also list the polar strength of the dipole component 
of the multipolar magnetic field of each star.

In Table \ref{params_fund} we also highlight which stars in our sample are part of binary systems and the binary separation.  For those with 
large separations the presence of a companion star is not expected to have any influence of over the stellar magnetic
field topology.  Those with the smallest separations, V2247~Oph and V4046~Sgr~AB, are found to host 
complex non-axisymmetric magnetic fields \citep{don10a,don11b}. We find little difference when comparing the field complexity 
found on single and binary stars.  For example, both components of HD~155555 \citep{dun08}, 
a tidally locked close post T Tauri binary, have magnetic field topologies and surface differential rotation measurements that are consistent with
those of single stars with similar spectral types \citep{dun08b}.  Similar results have also being found for the M-dwarf eclipsing binary 
YY~Gem (Morin et al., in prep.).  Therefore, we do not expect that binarity plays a significant role in setting the field 
complexity, although it clearly plays a role in the evolution of the large-scale coronal field due to the 
interaction between the stellar magnetospheres if the binary separation is sufficiently small, e.g. DQ~Tau \citep{sal10,get11}.  Such
large-scale changes in the coronal field that triggers flares appears to be generated by only small changes in the surface field 
topology, as determined by contemporaneous spectropolarimetric and X-ray observations \citep{hus07}.      


\section{Information from the Hertzsprung-Russell diagram}\label{hrdiagrams}
In this section we construct HR diagrams using the stars listed in Table \ref{params_fund} and contrast mass, age, and internal
structure properties derived from two different PMS evolutionary models.  We discuss the variation in the magnetic field topology
of stars across the diagram and explore the similarities between the field topologies of PMS stars and main sequence M-dwarfs with 
similar internal structures.


\subsection{Magnetic field topology and stellar internal structure}\label{new3.1}
\begin{figure}[!t]
  \centering
     \includegraphics[width=65mm]{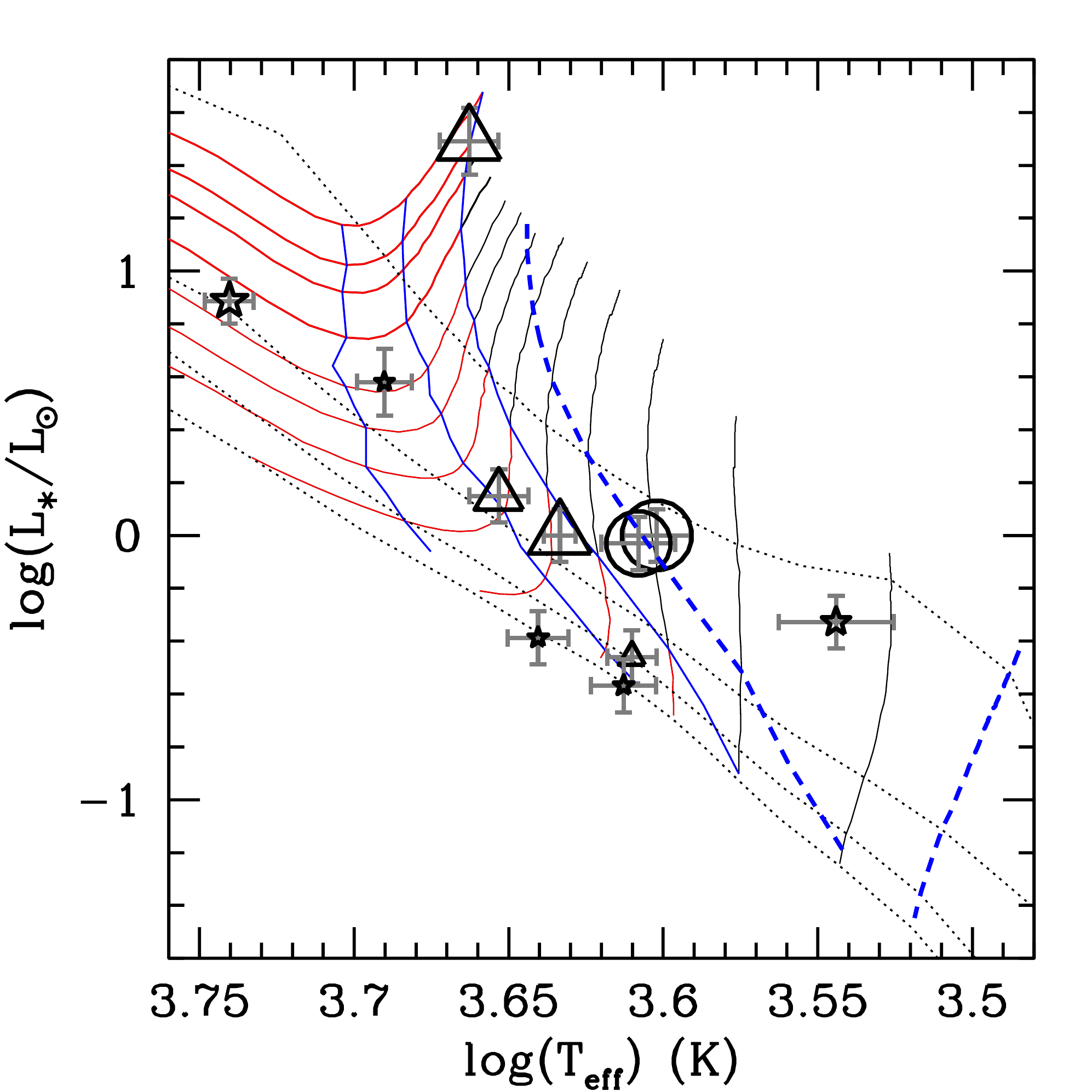}
     \includegraphics[width=65mm]{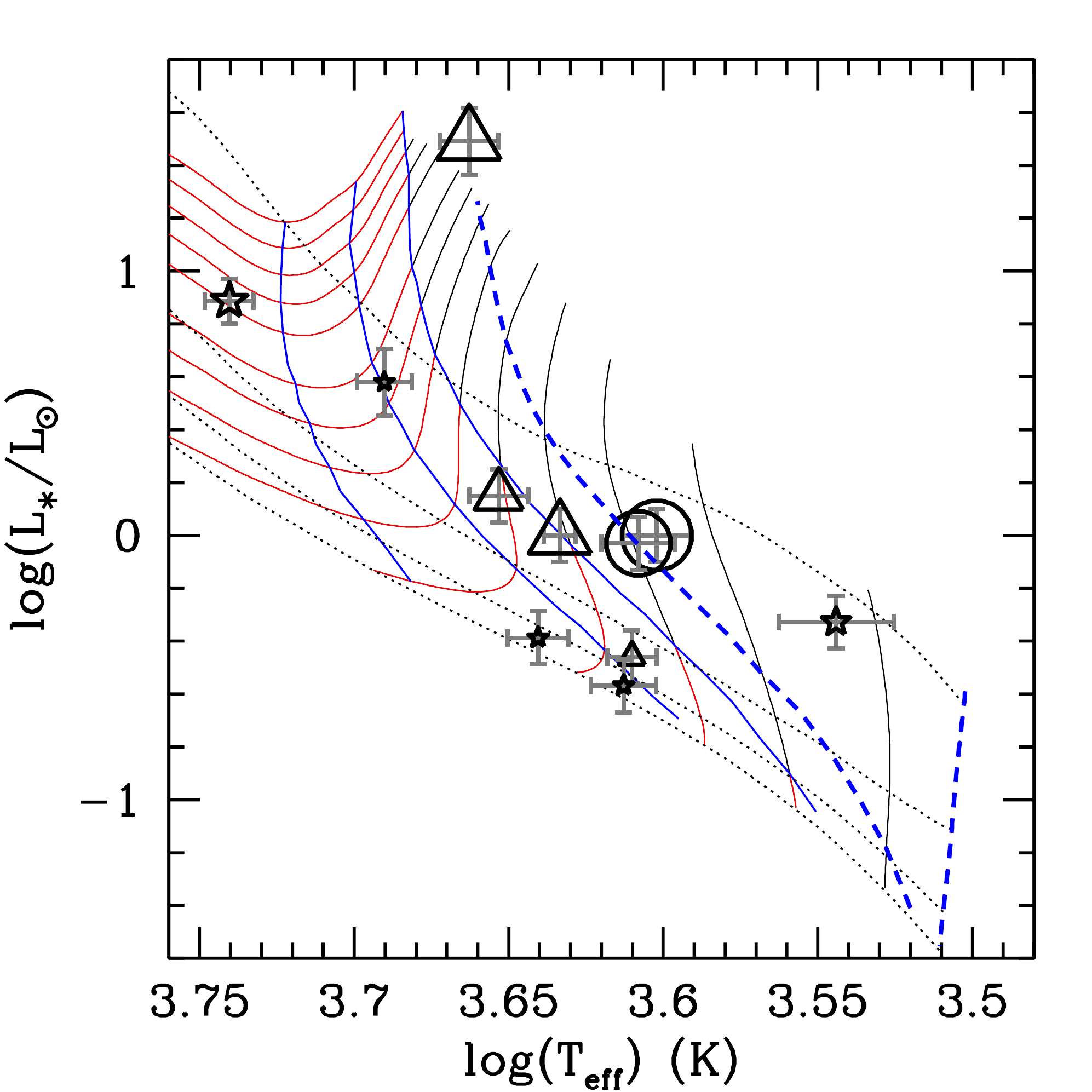}
  \caption{HR diagrams constructed from the \citet{sie00} PMS evolution models (left) and the Pisa models (right; \citealt{tog11}).  The 
               mass tracks (solid black/red lines) from right to left represent $M_\ast=0.3\,{\rm M}_\odot$ to $1.9\,{\rm M}_\odot$ in steps of
               $0.2\,{\rm M}_\odot$, then $2.2, 2.5, 2.7, 3.0\,{\rm M}_\odot$ for the \citet{sie00} diagram, and 
               $2.2, 2.4, 2.6, 2.8, 3.0\,{\rm M}_\odot$ for the \citet{tog11} diagram (due to the differing grid resolutions
               of the models). Black (red) segments represent stars with fully convective interiors (those which have developed radiative cores).  
               The solid blue lines are internal structure 
               contours, with the right-hand line the fully convective limit, the middle (left-hand) line is the loci of stars with core masses of 
               $M_{\rm core}/M_\ast=0.4$ (0.8).  Isochrones (dotted lines) are shown for ages of 
               $1, 5, 10, 15\,{\rm Myr}$ from upper right to lower left.   
               The black symbols are stars with published magnetic maps (see 
               Table \ref{params}), with circles stars with dominant dipole components and axisymmetric fields, triangles stars 
               with dominant high order field components
               ($\ell>1$) and axisymmetric magnetic fields, and asterisks stars with complex non-axisymmetric magnetic fields 
               with weak dipole components.  The
               size of the symbol is proportional to the stellar rotation period.  Stars which have spent longer with radiative cores (weaker dipole
               components) are typically faster rotators than fully convective stars (with strong dipole components) in the intermediate and 
               high mass regime ($\gtrsim0.5\,{\rm M}_\odot$).  The dashed blue lines are upper and lower limits on a boundary separating two
               different magnetic topology regimes within the fully convective region of the HR diagram; see \S\ref{lowmass}.
               }
  \label{hr}
\end{figure}

Figure \ref{hr} shows HR diagrams constructed from the \cite{sie00} and the Pisa \citep{tog11} PMS stellar evolution models.
The mass tracks are colored according to the internal structure of the star - black for fully convective stars, and red for partially
convective stars with radiative cores.  The HR diagrams also include internal
structure contours, the solid blue lines that connect stars of different effective temperature and luminosity, equivalently 
mass and age, but with the same internal structure (the same values of $M_{\rm core}/M_\ast$).   
The solid blue line on the right is the fully convective limit.  Stars that lie in the region of the HR 
diagram above and to the right of the fully convective limit have fully convective interiors, and those in the region below and to the left have 
radiative cores, or are entirely radiative for the more massive stars beyond a certain age.

The points in Figure \ref{hr} are the stars listed in Table \ref{params_fund} and Appendix \ref{starinfo} with different
symbols representing stars with different large scale magnetic field topologies, as detailed in the caption.  The stellar masses, ages, and radiative
core masses derived from the \citet{sie00} and the Pisa \citep{tog11} models are listed in Table \ref{params}.  Generally the masses derived from
both models agree to within $\sim$10\%, at least for the small sample of stars considered in this work.
The exceptions are V2129~Oph and MT~Ori which have masses
$\sim$20\% and $\sim$40\% larger respectively in the \citet{sie00} models.  With the exception of the lowest mass star
V2247~Oph, the isochronal ages are consistently younger in the \cite{tog11} models.  The largest age 
differences occur for high mass T Tauri stars, as well as for TW~Hya.  

It can be seen from Figure \ref{hr} that more massive PMS stars leave 
their Hayashi tracks at a younger age and their radiative cores grow
more rapidly than those of lower mass stars (see Appendix \ref{fullylimit} for further details).
There is no general trend in the core mass relative
to the stellar mass ($M_{\rm core}/M_\ast$) between the models; some stars have larger cores in one model, but in the same model
other stars have smaller cores.  Nevertheless it is encouraging that, with the exception of MT~Ori, if a given star has ended the fully
convective phase in one model it has also done so in the other model.  In this paper we choose to consider variations in the 
stellar internal structure by considering the fractional radiative core mass $M_{\rm core}/M_\ast$
rather than considering the variation in the fractional radiative core radius $R_{\rm core}/R_\ast$.  
This is because once a core develops a change in the ratio $M_{\rm core}/M_\ast$
is a direct reflection of the growth of the radiative core, assuming that in the T Tauri phase the star is no longer accumulating
significant mass via spherical infall and the stellar mass is set.  In contrast, changes in the ratio $R_{\rm core}/R_\ast$ represents both the growth
of the core, and the radius decrease of the contracting PMS star, see Figure \ref{age_core}.
The mass of the radiative core is therefore our preferred internal structure proxy, although it is directly related to the 
core radius as $M_{\rm core}\propto R_{\rm core}^3$ for a polytropic star.  Once a star evolves on to the main sequence
and its internal structure and radius have settled, both internal structure proxies can be used.

\subsection{Intermediate and high mass T Tauri stars ($\gtrsim$0.5$\,{\rm M}_\odot$)}\label{intermed}
It appears that the general characteristics of the large scale magnetic topology of an accreting T Tauri star ($\gtrsim$0.5$\,{\rm M}_\odot$) 
are strongly related to the star's position in the HR diagram (see Figure \ref{hr}). Stars which have similar internal structures (but 
very different mass/age and effective temperature/luminosity) appear to have similar magnetic field topologies: (i) stars in the 
completely convective regime (at least those above $\sim$0.5$\,{\rm M}_\odot$ at an age of 
$\sim$few ${\rm Myr}$; see \S\ref{lowmass} where we discuss low mass T Tauri stars) have strong dipole components to 
their magnetic fields and their fields are dominantly axisymmetric (AA~Tau and BP~Tau); (ii) stars with small radiative cores 
and large outer convective zones have magnetic fields that are dominantly axisymmetric and have high order 
components that dominate the dipole (V2129~Oph, GQ~Lup, TW~Hya, and MT~Ori - at least in the \citet{sie00} model for the latter, 
see below); and (iii) more evolved stars with substantial radiative cores and small outer convective 
zones have complex non-axisymmetric magnetic fields with weak dipole components (V4046~Sgr~AB, CR~Cha and CV~Cha).  In 
general the larger the radiative core the more complex the large scale magnetic field, and the weaker the dipole component 
(see Table \ref{params}).

We note that the magnetic field of MT~Ori is largely axisymmetric and the octupole, the dotriacontrapole, and the $\ell=7$ field mode
dominate the dipole.  Its magnetic field is more similar to those of V2129~Oph, TW~Hya, and GQ~Lup, all 
of which have small radiative cores, and very different to those of the fully convective stars AA~Tau and BP~Tau.
It thus seems likely that MT~Ori has developed a radiative core, and the \citet{sie00} models give a more 
accurate representation of the stellar structure in this region of the HR diagram (the \citealt{tog11} models suggest 
that MT~Ori is still fully convective).  This argument is further supported by the observed trends in the field topology with 
varying internal structure for MS M-dwarfs on either side of the fully convective divide, as we discuss in \S\ref{mdwarf_comp}.  
  

\subsection{Low mass T Tauri stars ($\lesssim$0.5$\,{\rm M}_\odot$) - dynamo bistability?}\label{lowmass}
The low mass T Tauri regime ($\lesssim0.5\,{\rm M}_\odot$) is, with the exception of 
V2247~Oph, an unexplored region of the HR diagram in terms of stellar magnetic field topologies.  
Intriguingly the field topology of V2247~Oph is complex and non-axisymmetric
with a weak dipole component, and resembles the fields of more massive T Tauri stars with substantial radiative cores 
rather than that of the more massive fully convective stars.  We therefore speculate that a bistable dynamo process with weak and 
strong field branches operates amongst the lowest mass fully convective PMS stars, similar to the lowest mass MS M-dwarfs 
discussed at the end of \S\ref{intro} (see \citealt{mor11}, their figure 4 in particular).  V2247~Oph would then belong to the 
weak field branch, while another fully convective star with similar stellar parameters but which belonged to the 
strong field dynamo branch, would host a simple magnetic field with a strong dipole component.  Once 
the low mass T Tauri regime has been explored in detail we expect that stars with a variety of field topologies will be found, 
some with weak dipole components corresponding to the weak field dynamo branch, and some with strong dipole components 
corresponding to the strong field dynamo branch.

As magnetic maps have yet to be obtained for the lowest mass fully convective PMS stars
the exact boundary between the strong dipole component regime and the bistable dynamo regime across the HR diagram
is unconstrained observationally, and also theoretically.  As bistable dynamo behavior for MS M-dwarfs only 
occurs for stellar masses $\lesssim0.2\,{\rm M}_\odot$ it is tempting to use this as the boundary separating the regions of
fully convective PMS stars with strong dipole components (those with $M_\ast\gtrsim0.2\,{\rm M}_\odot$) and those fully convective stars
where some host fields with strong dipole components while other stars with similar parameters host complex fields with
weak dipole components (those with $M_\ast\lesssim0.2\,{\rm M}_\odot$).\footnote{Recently \citet{sch12} have published a 
series of numerical simulations that suggest that bistable dynamo behavior can occur for all MS M-dwarfs in the fully convective 
regime, including those close to the fully convective limit.  Presently there is no observational evidence for this, with
bistable behavior only apparent for stars (both MS and PMS) located well below the fully convective limit.}  
The $0.2\,{\rm M}_\odot$ boundary is illustrated as the right-hand
dashed blue line in the HR diagrams in Figure \ref{hr}.  However, for MS M-dwarfs whose internal structure and therefore presumably the 
dynamo magnetic field generation process has settled, the $0.2\,{\rm M}_\odot$ boundary is $\sim$60\% of the 
MS fully convective limit of $\sim$0.35$\,{\rm M}_\odot$.  PMS stars are still contracting, however, and as the 
boundary between stars which are fully convective and those which are not is a function of age (see Appendix \ref{fullylimit}), we can also speculate that
the mass boundary below which bistable dynamo behavior occurs is itself a function of age.  The left-hand dashed blue line
in Figure \ref{hr} illustrates this alternative boundary which occurs at a stellar mass that is 60\% of the fully convective limit
at that age.  Thus, for a given bistable dynamo boundary, fully convective stars to the left of the boundary in the HR diagram would  
have simple axisymmetric fields with strong dipole components; stars to the right would be in the bistable regime and may
host a variety of field topologies just like the latest spectral type (lowest mass) M-dwarfs.  Taking the age dependent boundary defined in 
Figure \ref{hr} would mean that both AA~Tau and BP~Tau are actually in the bistable dynamo regime, and with 
simple magnetic fields with strong dipole components they would be on the strong field dynamo branch.  In reality the two 
dashed blue lines in Figure \ref{hr} likely represent upper and lower limits to the true bistable dynamo limit.  Clearly more data, and 
in particular more ZDI studies, are required for fully convective T Tauri stars to better constrain 
this limit and test our predictions.      


\subsection{Comparison with the magnetic topologies of MS M-dwarfs}\label{mdwarf_comp}
Although the links between T Tauri magnetic field topologies and stellar internal 
structure discussed in \S\ref{intermed} \& \ref{lowmass} are thus far based on a limited sample of PMS stars, 
similar trends have been found for MS M-dwarfs on either side of the fully convective divide \citep{don08a,mor08,mor10}.  
For MS M-dwarfs the transition from dominantly axisymmetric to non-axisymmetric 
fields occurs once the stellar mass exceeds $\sim$0.5$\,{\rm M}_\odot$ (see Figure \ref{mdwarfs}) which corresponds roughly to 
$M_{\rm core}/M_\ast\approx0.4$ (the \citealt{sie00} models give $M_{\rm core}/M_\ast=0.26$ for M-dwarfs of mass 0.5$\,{\rm M}_\odot$ and 
0.44 for 0.6$\,{\rm M}_\odot$).

The trends in magnetic topology of MS M-dwarfs across the fully convective limit thus roughly match those found from ZDI studies of 
PMS stars with similar internal structures, although there may be one subtle difference.  T Tauri stars with small radiative cores 
($0<M_{\rm core}/M_\ast\lesssim0.4$) host axisymmetric magnetic fields but field modes of higher order than the 
dipole dominate (typically, but not always, the octupole e.g. TW~Hya, V2129~Oph, and GQ~Lup).  In contrast, M-dwarfs with similar 
fractional radiative core mass ($M_{\rm core}/M_\ast$) host axisymmetric fields but the dipole component (although similarly weaker 
than that of fully convective stars) is the dominant field mode.  These M-dwarfs have small inclinations, closer to pole-on.  
In such cases it is difficult to recover field topology information at low stellar latitudes, and to reliably infer field modes above the dipole
when the dipole component is strong (large polar cool spots, e.g. that on TW Hya, help alleviate this problem in PMS stars with similarly low 
inclinations).  This apparent difference between the PMS and main sequence sample may just be observational bias.  

Alternatively, if the difference between the samples is real, it may be due to the rapidly changing internal 
structure of a PMS star with the core continuing to grow as the star evolves towards the main sequence.  The growth rate of the 
core (the rate of increase of the ratio $M_{\rm core}/M_\ast$) is more rapid the higher the stellar mass (see Appendix \ref{fullylimit}).  Thus as 
higher mass stars transition from fully convective to partially convective the dipole component of their magnetic fields may decay more
rapidly the faster the core develops.  Taking V2129~Oph and TW~Hya as examples, although both currently have small radiative cores similar
in size to mid spectral type M-dwarfs ($\sim$M4-M5 or 0.35-0.5$\,{\rm M}_\odot$) of $\sim$20\% of their stellar mass, by the time they arrive 
on the main sequence they will have substantial radiative cores with $M_{\rm core}/M_\ast\approx0.95$ and $\approx0.75$ respectively, and 
be of spectral type $\sim$F7 and $\sim$K3 \citep{sie00}.  Although TW~Hya and V2129~Oph currently have 
internal structures that are comparable to mid M-dwarfs they will differ substantially by the time they arrive on the main sequence.  By this
stage their field topologies will likely resemble the more complex fields found for stars with small outer convection 
zones, like CR~Cha, CV~Cha and V4046~Sgr~AB, and the earlier spectral type M-dwarfs.  In other words we are observing the PMS 
stars at a stage of their evolution where their large-scale magnetic fields are in the process of transitioning from simple to more complex 
fields.  This may explain the one subtle difference between PMS magnetic field topologies and those of MS M-dwarfs with currently 
similar internal structures. 


\section{Can we predict the magnetic field topology of T Tauri stars?}\label{predictHRD}

\subsection{The magnetic Hertzsprung-Russell diagram}\label{canwepredict}
\begin{figure}[!t]
  \centering
    \includegraphics[width=65mm]{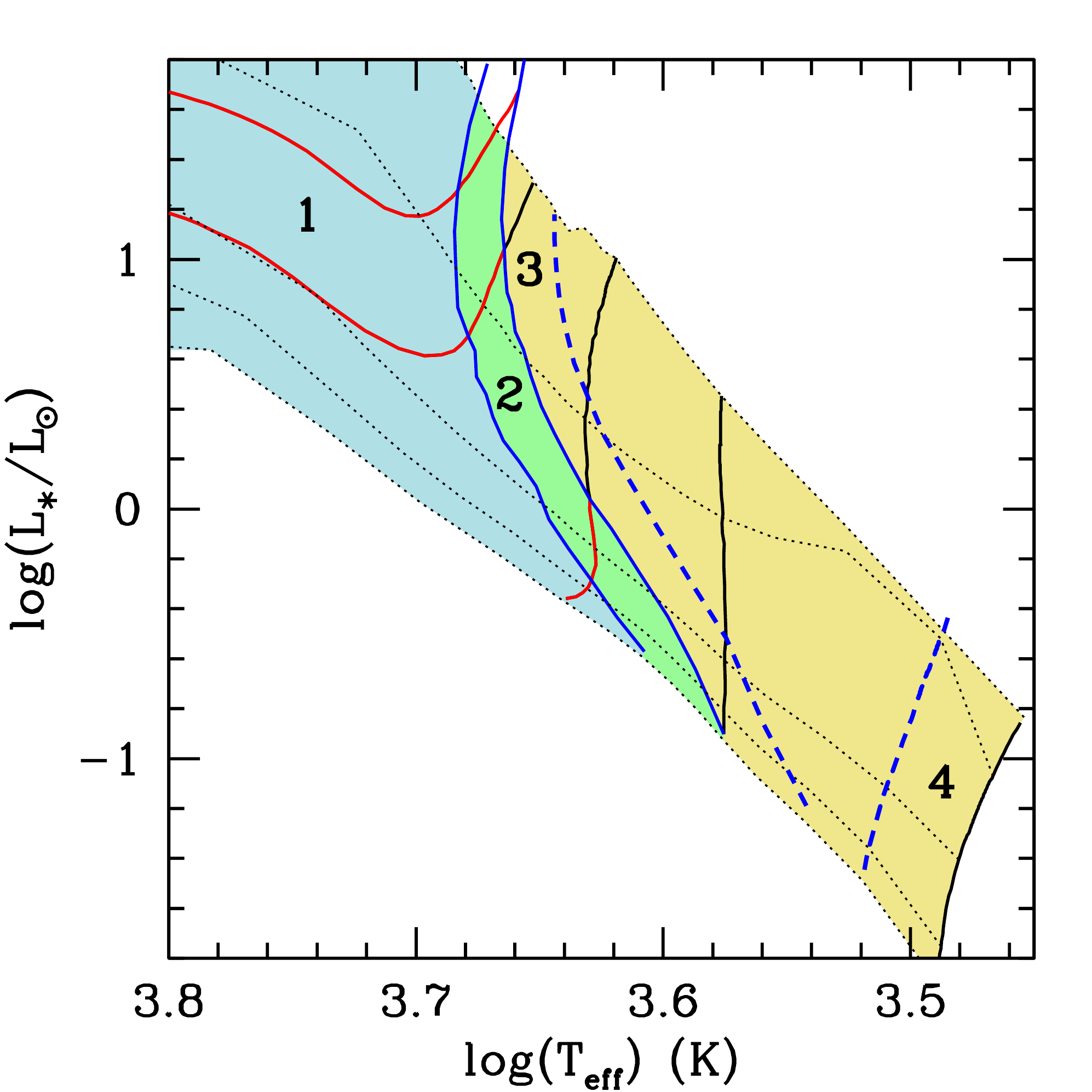}
  \caption{HR diagram constructed from the models of \citet{sie00} assuming $Z=0.02$ with convective overshooting.
                The mass tracks, for $M_\ast=0.1, 0.5, 1.0, 2.0$ and $3.0\,{\rm M}_\odot$, are black in the fully convective phase and
                red in the radiative core phase.  The isochrones (dotted lines) are for ages of 0.25, 1, 5, 10 and 15$\,{\rm Myr}$.  The solid 
                blue lines connect stars in the HR diagram with the same internal structure; on the right, it represents the fully convective
                limit, and on the left stars with $M_{\rm core}/M_\ast=0.4$.  ZDI studies of T Tauri stars, and the comparable magnetic trends 
                measured for main-sequence M-dwarfs with similar stellar internal structure, suggest that the general magnetic topology 
                characteristics of stars vary across the different colored regions.  In the blue region (region 1 - stars with substantial radiative cores; 
                $M_{\rm core}/M_\ast\gtrsim0.4$) stars have complex fields and weak dipole components.  In the green region 
                (region 2 - stars with small radiative cores; $0<M_{\rm core}/M_\ast\lesssim0.4$) stars have 
                magnetic fields that are largely axisymmetric
                with dominant high order field components ($\ell>1)$.  
                In the yellow region stars are fully convective.  Stars close to the fully convective limit (region 3) have strong 
                dipole components to their multipolar magnetic fields.  Below some boundary within the fully convective regime we 
                expect bistable dynamo behavior and to find a mixture of stars, some with simple 
                axisymmetric fields with strong dipole components and some with 
                complex non-axisymmetric fields with weak dipole components.
                The dashed blue lines, defined in \S\ref{lowmass}, denote possible upper and lower limits between this, region 4, and region 3.  
                We stress that the region boundaries are empirical and are 
                poorly constrained observationally and theoretically (see \S\ref{limitations}).  More data is required to confirm the exact 
                boundaries which may also vary with stellar mass.          
               }
  \label{color_hr}
\end{figure}

As summarized above, ZDI studies have revealed that T Tauri stars host multipolar magnetic fields; however the field topology
seems to be strongly linked to the stellar internal structure, and consequently to how the magnetic field is generated and maintained by
differing dynamo mechanisms. Empirically we define four distinct magnetic topology regions across the PMS,
see Figure \ref{color_hr}, defined as we move from upper-left (warm/luminous) to lower-right (cool/faint) in the HR diagram:
\begin{itemize} 
\item{Region 1 (blue in Figure \ref{color_hr}): stars with substantial radiative cores 
$M_{\rm core}/M_\ast\gtrsim0.4$.  In this region stars have complex 
magnetic fields with many high order components.  The fields are highly non-axisymmetric and the dipole component is weak. 
This region contains the most massive T Tauri stars, typically those of spectral type G or early K, and also older stars of later spectral type.
V4046~Sgr~AB, CR~Cha and CV~Cha lie in this region.}
\item{Region 2 (green in Figure \ref{color_hr}): stars with small radiative cores $0<M_{\rm core}/M_\ast\lesssim0.4$. In 
this region stars have magnetic fields that are dominated by strong high order field components.  The dipole component 
may be weak or strong but it contains less magnetic energy than the higher order field modes.  The fields are largely 
axisymmetric.  MT~Ori, TW~Hya, V2129~Oph, and GQ~Lup, lie in this region.}
\item{Region 3 (yellow in Figure \ref{color_hr}): fully convective stars to the left of some boundary between the dashed 
blue lines. In this region stellar magnetic fields are axisymmetric with strong (kilo-Gauss) dipole 
components.  AA~Tau and BP~Tau likely lie in this region.}
\item{Region 4 (within the yellow region in Figure \ref{color_hr}): fully convective stars to the right of some 
boundary between the dashed blue lines.  The boundary between this region and region 3 is not well defined observationally.  
The dashed blue lines in Figure \ref{color_hr} are possible upper and lower limits to the true boundary
(see \S\ref{lowmass}).  By comparison to the magnetic topologies of the lowest mass 
fully convective M-dwarfs we expect that this region will be populated by stars 
with a mix of magnetic topologies, the dynamo process being bistable with a strong and weak 
field branch \citep{mor11}.  Stars on the strong (weak) field branch will have fields similar to stars in region 3 (1). 
V2247~Oph lies in this region and on the weak field dynamo branch.}
\end{itemize} 

The general magnetic topology characteristics of a given T Tauri star will change with age as the star
evolves down its mass track towards the MS.  
In the high and intermediate mass regime, stars with mass
$\gtrsim0.5\,{\rm M}_\odot$, T Tauri stars initially host magnetic fields that are axisymmetric with a strong
dipole component.  As the fully convective phase of evolution ends and a small radiative core develops the field remains 
largely axisymmetric but the dipole component decays away, leaving a field that is dominated by strong high order field 
components (those with $\ell>1$).  By the time that the core mass exceeds $M_{\rm core}/M_\ast\approx0.4$ the dipole component is weak, 
and the field is complex having lost its earlier axisymmetry.  This core mass boundary is empirical, based on the limited sample of stars where
magnetic maps have been published to date, and is therefore somewhat speculative at this stage.
Clearly more data is required to confirm the exact value, but we note that currently unpublished data is consistent with these trends.   
The boundary may be more fluid, and itself dependent on stellar
mass given that for higher mass stars the growth rate of the radiative core is more 
rapid than for lower mass stars (see Appendix \ref{fullylimit}).  Furthermore, 
if the boundary between fully convective stars with simple fields and those in the bistable regime is as extreme as masses below $60$\% 
of the fully convective limit (see the discussion in \S\ref{lowmass}; and the left-hand blue dashed line in Figure \ref{color_hr}) then this 
picture would have to be modified, as some stars would be born within the bistable regime and host fields with weak dipole components 
which would then strengthen as the stars approach the fully convective limit.  It is not clear what interplay between the stellar contraction,
rotation period and mass could influence the dynamo process in this way.   

The general T Tauri magnetic topology trends are currently empirical and are based
on knowledge garnered from observationally derived magnetic maps.  However, a magnetic topology change 
due to the transition from a fully to a partially convective stellar interior is further supported by the observed changes in periodic 
variability and X-ray luminosities \citep{reb06,sau09,may10}.  If ZDI data acquired in future continues to follow
the empirical trends, then in principle it is possible to infer the general properties of a T Tauri 
star's large scale magnetic field solely from its position in the HR diagram.  

We do not claim that it is possible to 
know the exact properties of a star's magnetic field based solely on its effective temperature and luminosity.  Indeed
we expect that the large scale field topology, and the strength of the various magnetic field components, will evolve in time due to 
magnetic cycles (see \citealt{don11a,don11c,don12} for 
discussion about the changes in the fields on V2129~Oph, TW~Hya, and GQ~Lup, although longer timescale observing programs 
potentially spanning several years are required to search for and confirm the existence of magnetic cycles on T Tauri stars).
Nonetheless, it appears as though it is possible to estimate the general properties of the stellar magnetosphere, 
for example whether or not the field will be dominantly axisymmetric with a weak dipole component, or axisymmetric with strong 
higher order components, or if the field will be highly complex with many multipolar 
components and non-axisymmetric.  Our ability to do so, however, depends on both the accuracy with 
which the star has been positioned in the HR diagram and on the veracity of the PMS stellar evolution models.


\subsection{Limitations: observational \& theoretical}\label{limitations}
Our ability to ascertain the magnetic topology of a given star from the HR diagram would be limited by how well we can position the star in the
HR diagram and on the dependability of the PMS evolution models (see \citealt{hil08} for a detailed review).  
Observationally the challenges lie in the assignment of a stellar spectral type, and the subsequent conversion to 
effective temperature with the assumption of some metallicity and surface gravity dependent scale.  Likewise
to discern the stellar luminosity we must carefully account for extinction; the presence of large surface cool spots
and the related photometric variability; for accreting PMS stars the additional luminosity from accretion;
uncertainties in the distance estimate; and for some sources the existence of unresolved close companions \citep{har01}.

Theoretically the errors that can arise from assumptions in the constituent input physics of the PMS 
evolution models have been succinctly summarized by \citet{sie01}, \citet{pal01} and \citet{tog11}, with the 
latter paper providing detailed comparison between different evolutionary models.  Modeling the evolution of
a forming star along its mass track and across the HR diagram is a formidable task.  Errors, as well as differences between the various
available PMS evolution models, arise from differing assumptions about the equation of state; the adopted 
boundary conditions, for example whether a grey or more realistic atmosphere model is employed; how convection is
handled; the assumed metallicity; the effects of mass accretion and rotation; and the influence of different formation
histories during the protostellar phase.  Taking these effects into account, \citet{sie01} estimates the errors in the  
mass tracks to be $\Delta T_{\rm eff}\sim100-200\,{\rm K}$ and $\Delta\log{(L_\ast/{\rm L}_\odot)}\sim0.1$.  Additionally if 
magnetic fields themselves are not accounted for in models of convection, a further source of error is introduced 
to the models \citep{dan00}. Finally, most models do not account for episodic accretion which may alter the stellar structure and the age at
which stars of a given mass develop a radiative core \citep{bar10}.

The uncertainty in the stellar evolution models themselves, and the observational difficulty in accurately assigning
effective temperatures and luminosities, must be kept in mind when using Figure \ref{color_hr} to predict the general
magnetic field properties of a particular T Tauri star.  However, turning the problem around, rather than using the star's
position in the HR diagram to ascertain its magnetic topology, it may be possible to use the observationally 
derived magnetic topology to test the accuracy of certain aspects of the PMS evolution models themselves.  
Just as dynamical mass measurements for binary stars \citep{pals01,hil04} and for single 
stars with disks \citep{sim00} can be used to constrain the accuracy of mass tracks, and the amount of lithium depletion   
isochronal ages (e.g. \citealt{pall05}; \citealt{sod10}), magnetic field topologies may be used to test stellar internal
structure information.  For example, in \S\ref{canwepredict} we argued that the field topology of T Tauri stars varies from 
simple and axisymmetric with a strong dipole component to complex and highly non-axisymmetric with a weak dipole component
with the growth of a radiative core.  A difference in the external field topology is expected given the different dynamo
process operating in fully convective stars compared to more evolved and/or more massive stars with outer convection zones, 
radiative cores, and stellar analogs of the solar tacholine.  The right hand solid blue line in 
Figure \ref{hr} denotes the fully convective limit.  By carrying out 
ZDI studies for stars around this limit, stark variation in the field topology between various stars may be used as a probe
to observationally constrain which regions of the HR diagram are populated by fully convective stars,  and which regions are populated
by stars with radiative cores.  In other words, by determining the regions of the HR diagram where stars with simple and complex magnetic fields
lie, we can determine whether or not the internal structure information derived from the PMS evolution models is accurate.


\section{Discussion}\label{discussion}

\subsection{The dipole component of T Tauri magnetospheres and the star-disk interaction}\label{dipcomp}
For accreting T Tauri stars it is generally the strength of the dipole component that is the most significant
in terms of controlling the disk truncation radius, even when the dipole component is weak 
compared to the higher order components \citep{ada11}.  This can be seen by considering the field strength at 
the inner disk truncation radius.  Let's consider a simple example of a star with a dipole plus 
an octupole field component, as many accreting T Tauri stars host large scale magnetic fields of this 
form \citep{gredon11}.  In the equatorial plane the contribution to
the vertical component of $\mathbf{B}$ threading the disk at a distance $r$ from the stellar center from the dipole component of polar strength 
$B_{\rm dip}$ is $B_{\rm dip}(R_\ast/r)^3/2$.  For the octupole field component
the equivalent expression is $3B_{\rm oct}(R_\ast/r)^5/8$ where $B_{\rm oct}$ is the polar strength of the octupole \citep{gre10}.
Thus, assuming a typical disk truncation radius of $5\,R_\ast$ (which is $\sim$70\% of the equatorial corotation radius for a $2\,{\rm R}_\odot$ 
solar mass star with a rotation period of $6\,{\rm d}$), then the ratio of the strength of the octupole to the dipole
component at the inner disk edge is $(3/100)(B_{\rm oct}/B_{\rm dip})$.  Taking the ratio of the polar strength of
the field components as $B_{\rm oct}/B_{\rm dip}=10$, which is larger than thus far observed for any T Tauri star (which
is $B_{\rm oct}/B_{\rm dip}\approx6$ for TW~Hya in March 2008; \citealt{don11c}) then the contribution to the field at the inner disk from
the octupole component is only 30$\%$ that of the dipole component, and becomes less significant the weaker 
(stronger) the octupole (dipole) component and for larger disk truncation radii\footnote{This simple illustrative example ignores the tilt of 
the multipole components which should be accounted for when calculating the field strength at the inner disk, and consequently the disk 
truncation radius \citep{gre08}.}.  The dipole is, in the majority of cases, the most significant field component in controlling the disk 
truncation radius. 

\begin{figure}[!t]
 \centering
   \includegraphics[width=65mm]{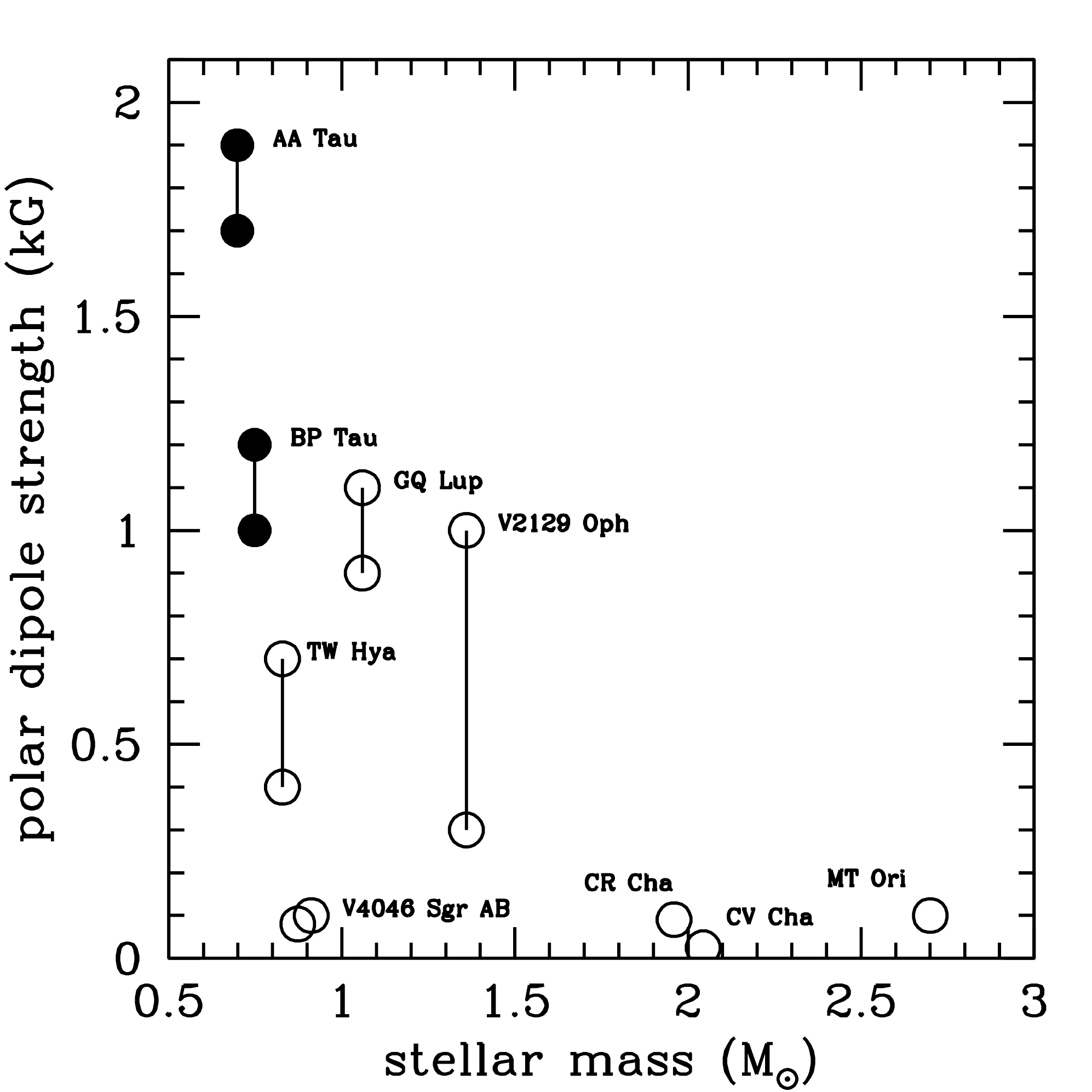}
   \includegraphics[width=65mm]{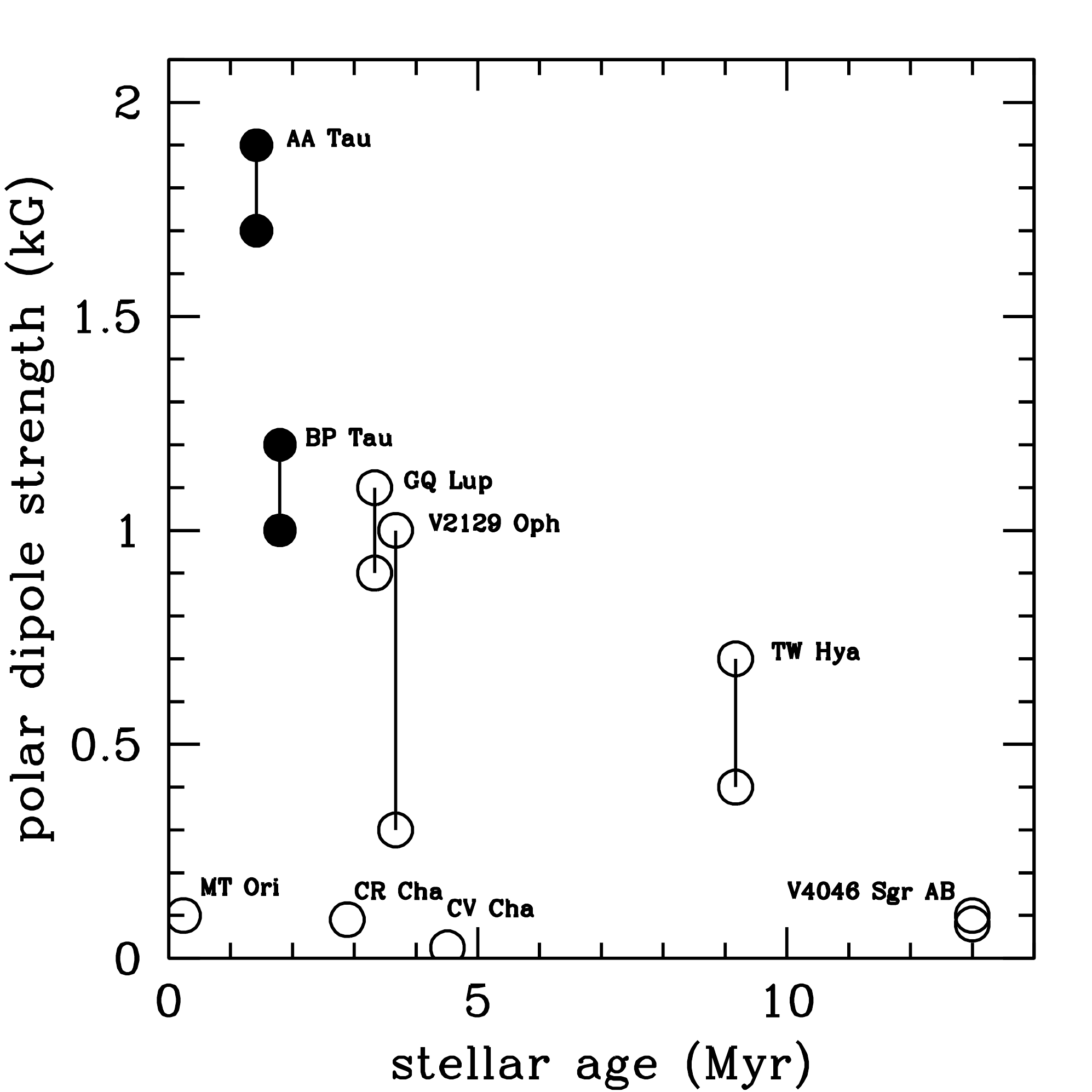}
 \caption{The polar strength of the dipole component of the magnetic fields of various accreting T Tauri stars as a function of
              stellar mass (left) and age (right) in the intermediate and high mass range ($\gtrsim$0.5$\,{\rm M}_\odot$) listed in
	      Table \ref{params}.  Filled circles denote fully convective stars, and open circles stars with radiative cores, as determined from the
               PMS evolution models of \citet{sie00}. The solid vertical lines join the same star observed at two different epochs.  It is clear that 
               higher mass and/or older stars have weaker dipole components than lower mass and/or younger stars.}
 \label{bdip_ctts}
\end{figure}

Figure \ref{bdip_ctts} shows the variation in the polar strength of the dipole component for high and intermediate 
mass T Tauri stars listed in Table \ref{params}.  It is clear that more massive and/or older T Tauri stars, those which
have ended the fully convective phase of evolution, have weaker dipole components than younger and/or lower mass stars.  
A possible exception is for some of the low mass T Tauri stars which may show a variety of field topologies (see \S\ref{lowmass}).  

The observed rapid decay in the dipole component with the growth of a radiative core can influence 
the star-disk interaction only if the fully convective phase ends before the disk has dispersed.  
As demonstrated in Appendix \ref{fullylimit} the age at which a radiative core develops is highly dependent on stellar mass, with high mass
T Tauri stars ($\gtrsim1.0\,{\rm M}_\odot$) ending the fully convective phase in $\lesssim2.6\,{\rm Myr}$ based on the \citet{sie00} models,
or $\lesssim2.2\,{\rm Myr}$ based on the \citet{tog11} models.  This timescale drops to as little as $0.5\,{\rm Myr}$ in both models 
for stars of $2\,{\rm M}_\odot$.  Therefore the drop in the dipole component, and the subsequent effect on the star-disk interaction, is more 
relevant for higher mass T Tauri stars than for lower mass stars as most of the latter will have lost their disks before the end of the 
fully convective phase.  However, there is also observational evidence that the disk lifetime is mass dependent with high mass
stars losing their disks faster than stars in the intermediate and low mass range \citep{car06,cur09,wil11}\footnote{Recent theoretical models 
suggest only a weak stellar mass dependence on disk lifetimes \citep{gor09,erc11}. Furthermore disk lifetimes may be influenced by the 
star forming environment \citep{luh08}.}.  
Therefore, the effect of the evolution of the large scale stellar magnetic field topology becomes a question of timescales.  For a given star
does the radiative core develop before it stops interacting with its circumstellar disk?  There are a number
of well studied T Tauri stars in the high and intermediate mass range which have developed radiative cores and 
which show evidence for significant ongoing accretion and substantial disks, for example the stars discussed in 
this paper as well as those studied by \citet{cal04}.  

In principle the rapid drop in the dipole component, and therefore
the field strength at the inner disk, at the end of fully convective phase will allow the disk to push closer to the star.  
This would lead to a increased spin-up torque acting on the star due to the magnetic links with the disk interior to the 
equatorial corotation radius that are rotating faster than the star (in addition to the spin up torques from accretion and 
the stellar contraction; \citealt{mat05}) and consequently an increase in the stellar rotation rate.   In contrast, fully convective
stars (at least those that are not in the weak field bistable dynamo regime) with strong dipole components should 
be able to maintain their slow rotation by truncating their disks out to, and perhaps even beyond, the 
corotation radius (the propeller regime; \citealt{rom04,don10b}).   
  
There is tentative evidence for this rotational evolution scenario within the sample of stars considered in this paper.  
In Figure \ref{hr} the size of the symbols is proportional to the rotation period of the star (listed in Table \ref{params}).
For intermediate and high mass T Tauri stars ($>$0.5$\,{\rm M}_\odot$) the fully convective stars which have strong 
dipole components are more slowly rotating than those which have ended the fully convective phase.
Additionally, stars which have spent longer with radiative cores (and therefore with weak dipole 
components) are, on average, rotating faster than the fully convective stars.  This strongly hints 
that the effect of the change in the magnetic topology with the development of a radiative core is that a PMS star enters a 
spin-up phase, if they are still interacting with their disks when this transition occurs.  However, this picture may be too simplistic, 
as the disk truncation radius is sensitive to parameters other than the magnetic field strength which will vary with time, and 
disk lifetimes are likely also a function of many parameters.  

The disk truncation radius depends on the stellar radius, the mass accretion rate, and the 
polar strength of the stellar dipole component $R_t \propto B_{\rm dip}^{4/7}R_\ast^{12/7}\dot{M}^{-2/7}$ 
(e.g. \citealt{kon91})\footnote{There is a difference in $R_t$ for multipolar compared to dipolar magnetospheres, although in most cases  
this difference is small (see section 6 of \citealt{ada11}).  Provided that the dipole component is not significantly tilted with respect to the stellar
rotation axis $R_t$ values can be estimated using the strength of the dipole component at the stellar rotation pole.  
The higher order field components, however, must be accounted for in models of accretion flow onto the star \citep{gredon11,ada11}.}.
Although a drop in the dipole component will allow the disk to push closer to the star (as will the reduction in the stellar radius, since
PMS stars are contracting) the observed drop in mass accretion rate with increasing stellar age (e.g. \citealt{har98,sic04}) 
has the opposite effect and allows the stellar magnetosphere to keep the disk at bay at a larger radius.  
Thus the disk truncation radius, and consequently the balance of torques in the star-disk system, will depend on the interplay 
between the rate of decay of the dipole component and the drop in the mass accretion rate with time.  Additionally magnetic 
cycles (the beginnings of which may have already been observed in V2129~Oph and GQ~Lup, see Appendix \ref{starinfo}) will cause
variations in the large scale field topology, and therefore the disk truncation radius, over time.  Thus a new generation of 
magnetospheric accretion models that track the rotational evolution of the star incorporating the time evolution
of the mass accretion rate, the stellar contraction (similar to those of \citealt{mat10,mat12}) and for the first time
the time evolution of magnetic fields following the observational correlations and potential magnetic cycles, are now warranted. 


\subsection{Magnetic field topology, rotation rate, and Rossby number}\label{tau_c}
Throughout this paper we have concentrated on the links between the stellar mass and age, which considered in tandem 
reveal the stellar internal structure, and the large-scale magnetic field topology.  However, the stellar rotation rate and 
the Rossby number, the ratio of the rotation period to the local convective turnover time in the stellar interior 
($Ro=P_{\rm rot}/\tau_{\rm c}$), may also influence the magnetic topology.  In order to search for such trends         
we require estimates of the convective turnover time $\tau_{\rm c}$.  As PMS stars are contracting, and especially 
once a radiative core begins to grow at the expense of the convective zone depth, $\tau_{\rm c}$
values are highly sensitive to the stellar mass and age.  Unfortunately most published $\tau_{\rm c}$ estimates come 
from models that track the stellar evolution over timescales of order Gyr and across a limited range of stellar mass \citep{gil86,kim96,lan10},
and lack the time and mass resolution required for our purposes.  An exception is the model of \citet{jun07}.  Y.-C. Kim has kindly
supplied us with finer resolution grids than published.  Our rough convective turnover time estimates are listed in Table \ref{params_fund}.  
The $\tau_{\rm c}$ values are calculated at a distance of half of the mixing length above the base of the convective zone.  Although it is
only an assumption that this is the depth where the dynamo operates, the same assumption is made
in the cited models and is fully consistent with the work of others (e.g. \citealt{joh00,pre05,ale12}).

\begin{figure}[!t]
 \centering
   \includegraphics[width=65mm]{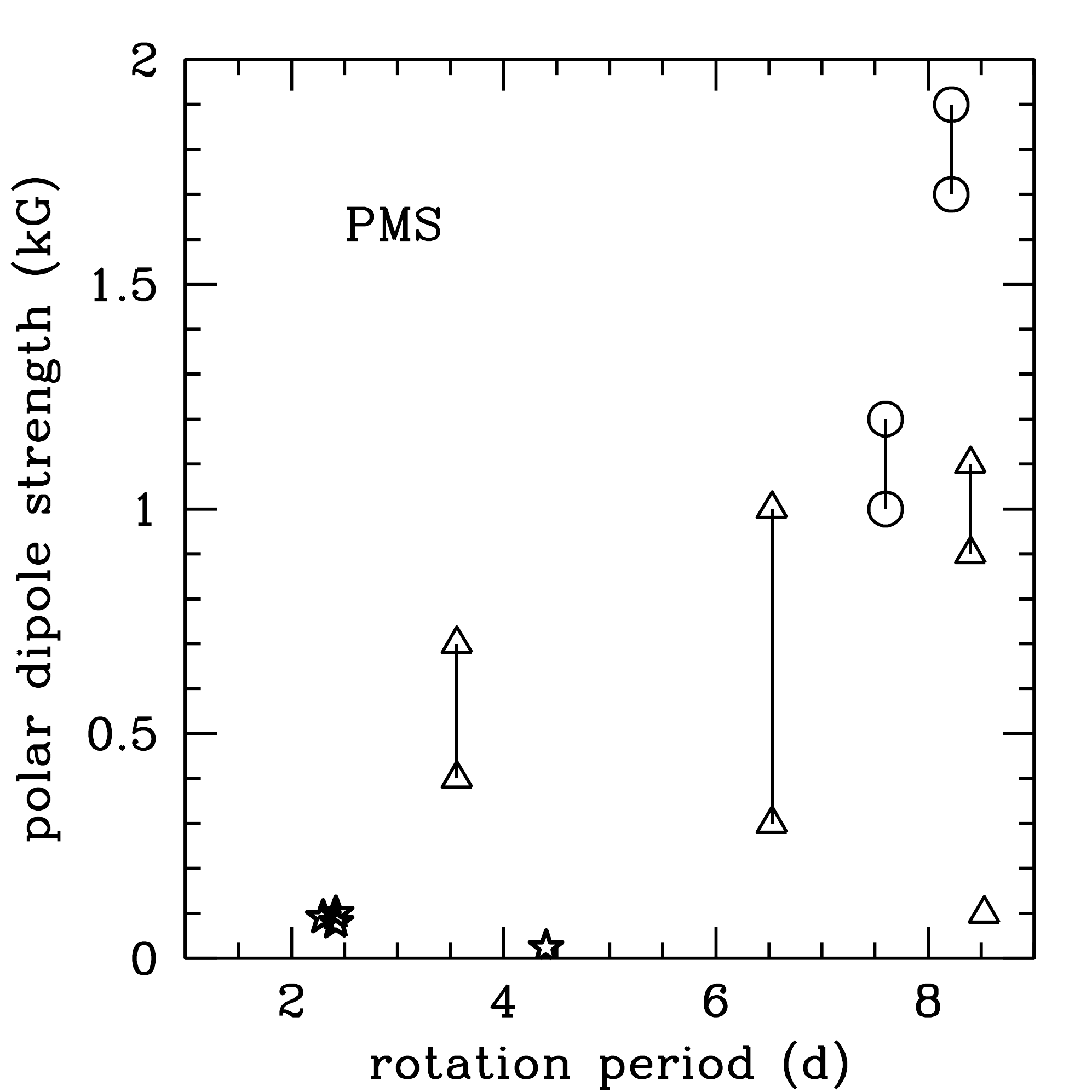}
   \includegraphics[width=65mm]{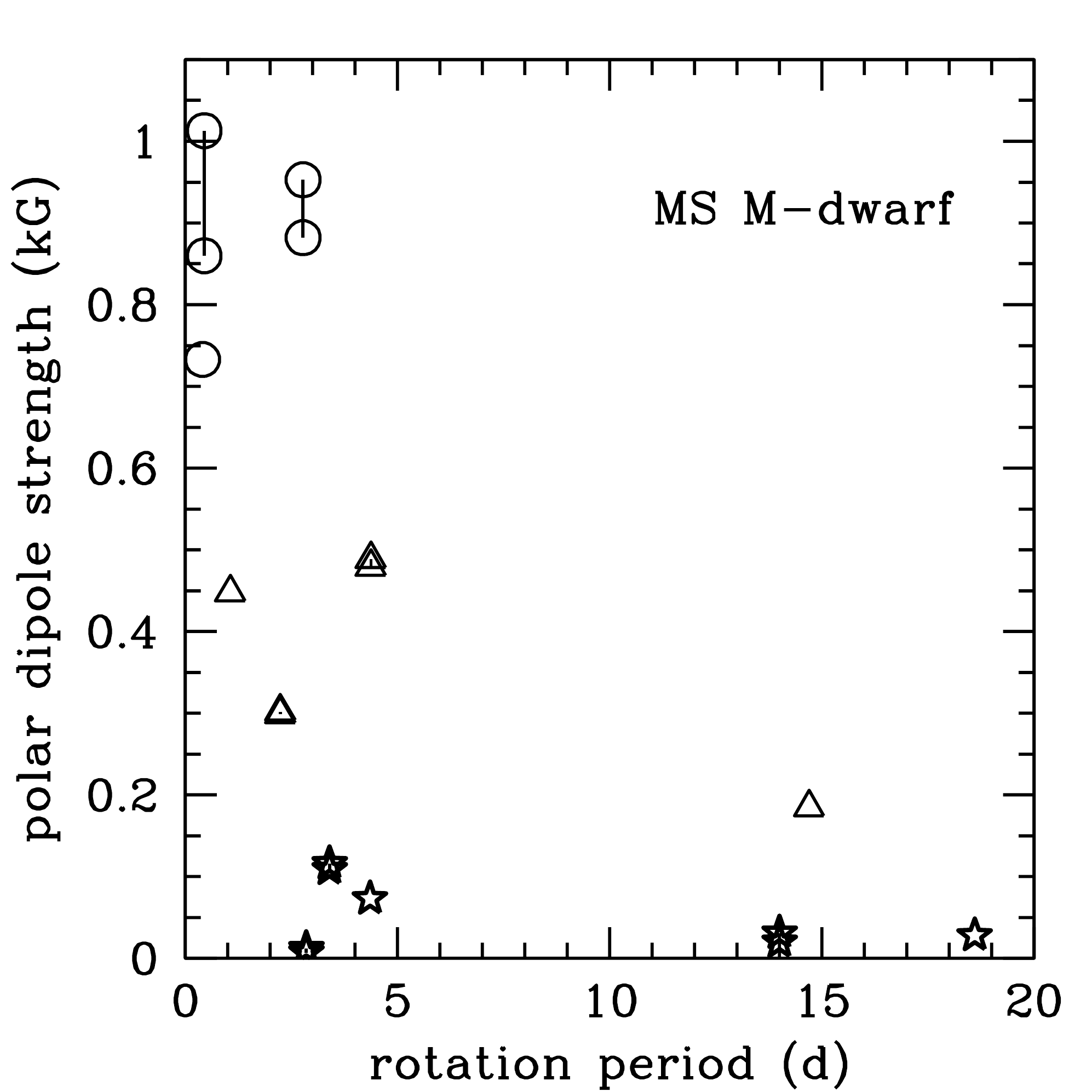}
    \includegraphics[width=65mm]{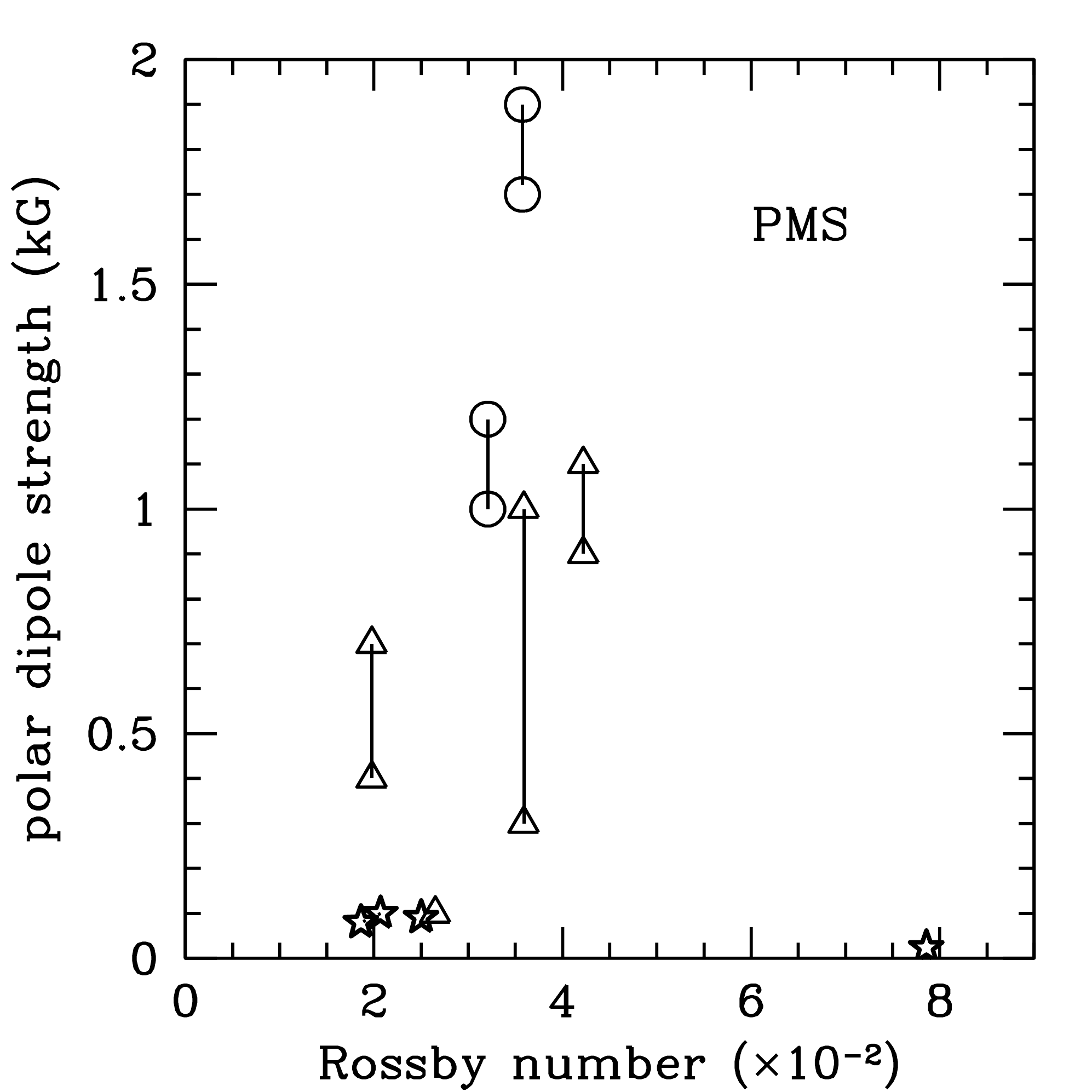}
    \includegraphics[width=65mm]{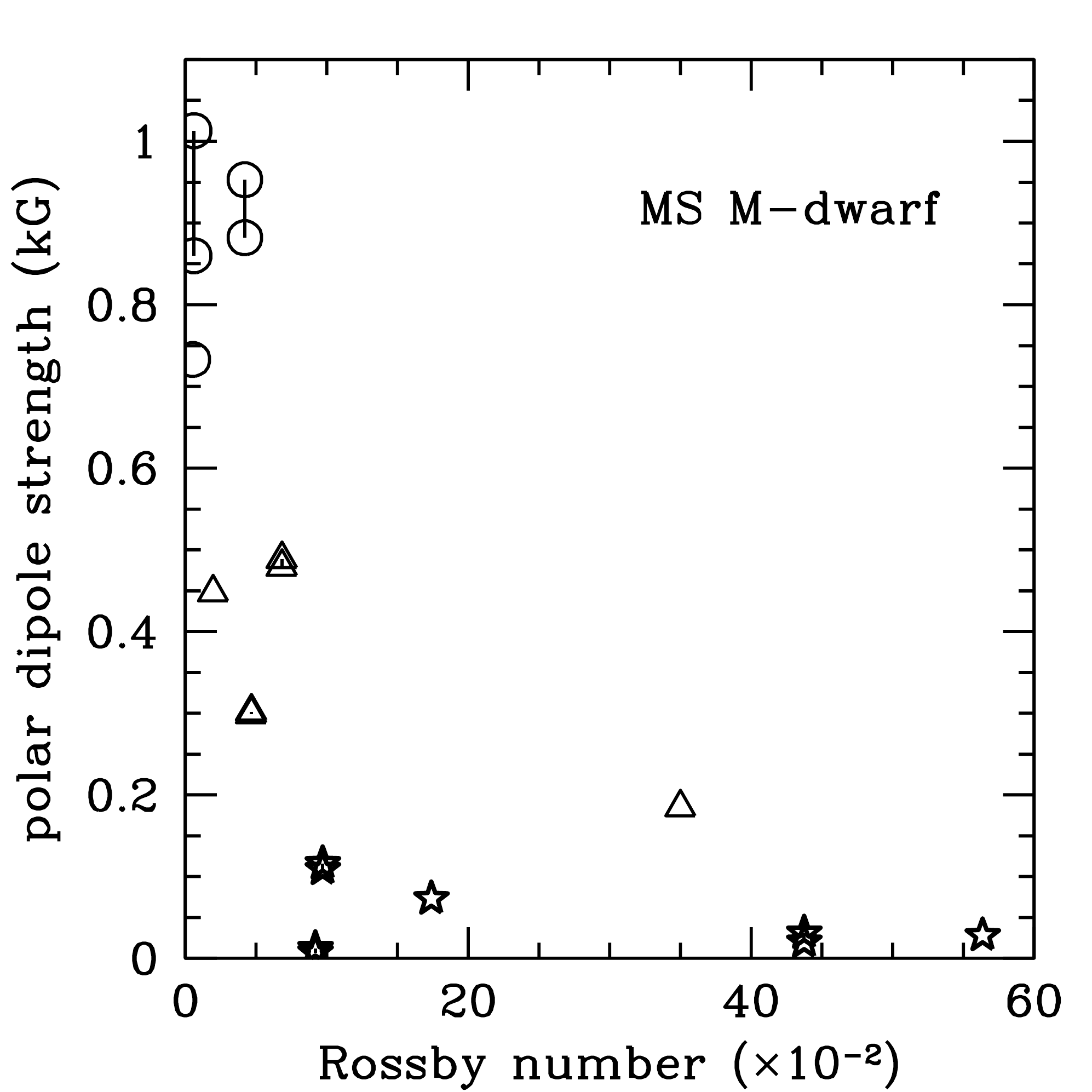}
 \caption{The polar strength of the dipole component of the multipolar magnetic fields of the intermediate/high mass PMS stars (left)
and MS M-dwarfs (right; early/mid-M spectral types from \citealt{don08a} and \citealt{mor08}) versus rotation
period (upper) and Rossby number (lower).  Points connected by vertical lines are stars observed at two epochs.  
Circles are fully convective stars, triangles (asterisks) stars with small (large) radiative
cores.  
}
 \label{magrot}
\end{figure}

Zeeman broadening studies (e.g. \citealt{joh07}) have found no links between the mean surface magnetic field strengths of 
PMS stars and rotation parameters.  This is perhaps not surprising given that all T Tauri stars lie in the saturated regime (with some 
into the supersaturated regime) of the well-defined MS rotation-activity relation: plots of the ratio of X-ray to bolometric luminosity versus 
Rossby number \citep{pre05}.  Zeeman broadening, which probes 
all of the small scale magnetic field regions close to the star (the tangling and reconnection of which gives rise to the X-ray emission), 
does not give access to information about the large-scale field topology.  In this work we are particularly interested in links between the
rotation parameters and the polar strength of the dipole component, given its importance to the star-disk interaction (see \S\ref{dipcomp}).

\begin{figure}[!t]
 \centering
   \includegraphics[width=65mm]{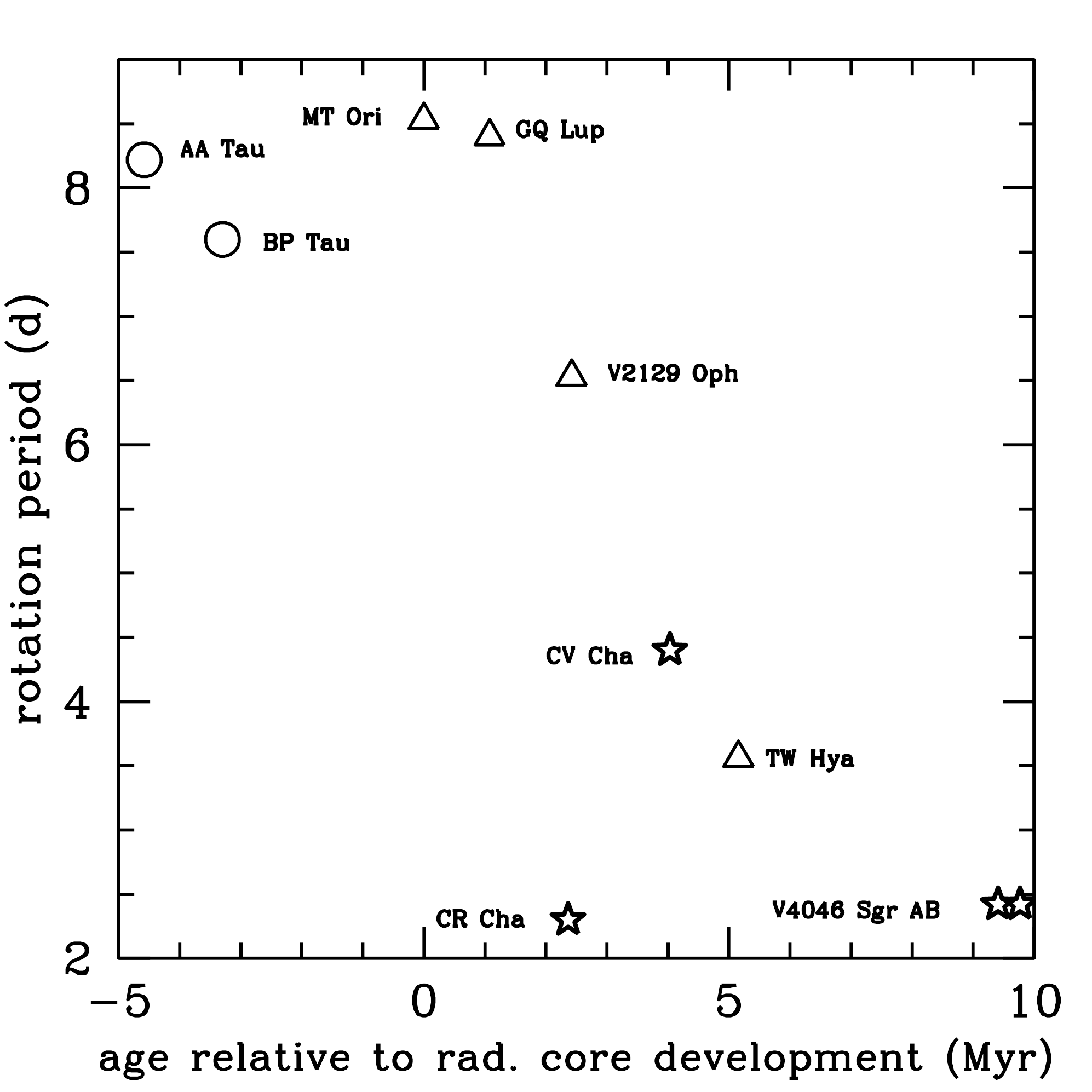}
 \caption{Stellar rotation periods for the intermediate/high mass accreting PMS stars as a function of time since (positive abscissa values)
              or time until (negative abscissa values) the development of a radiative core.  The symbols are as in Figure \ref{magrot}.  Stars which 
              have spent longer with radiative cores (and therefore with weaker dipole components, see Figure \ref{bdip_ctts}) are faster
              rotators.
	      }
 \label{spinup}
\end{figure} 

In Figure \ref{magrot} we present plots of $B_{\rm dip}$ versus the rotation parameters ($P_{\rm rot}$ and $Ro$; plots with $v_\ast\sin{i}$ 
as the abscissa, which are not shown, are similar to the $P_{\rm rot}$ plots) for both intermediate/high
mass PMS stars and the early/mid-spectral type MS M-dwarfs from \citet{mor08} and \citet{don08a}.  The 
M-dwarf sample spans the unsaturated and saturated regimes of the rotation-activity relation \citep{don08a}.  The saturated 
regime occurs at rotation periods of $P_{\rm rot}\lesssim4\,{\rm d}$ for M-dwarfs of mass $\sim$0.5$\,{\rm M}_\odot$, or 
at $Ro\lesssim0.1$ \citep{piz03}.  M-dwarfs which lie in the saturated regime (like all PMS stars) show little relation 
between $P_{\rm rot}$ and $B_{\rm dip}$ with 
a range of polar dipole strengths, being strongest for the fully convective stars (lowest mass; circles in Figure \ref{magrot} upper right panel), 
weakest for the stars with small outer convective zones (highest mass; asterisks), and of intermediate strength for stars slightly more massive than 
the MS fully convective limit (triangles), as already discussed - see Figure \ref{bdip_mdwarf}.\footnote{\citet{moriau}, their figure 2, 
and \citet{doniau}, his figure 1, provide more complete overviews of the links between mass, rotation period, field topology, and Rossby 
number for MS stars.  Such plots of the $M_\ast-P_{\rm rot}$ plane with contours of constant $Ro$ are not meaningful for our small 
sample of PMS stars, which span a range of stellar ages, as the $Ro$ contours are highly sensitive to age due to the variation 
of $\tau_{\rm c}$ values with the stellar contraction and radiative core growth.}  There does appear to be correlations between 
$B_{\rm dip}$ and $P_{\rm rot}$, and $B_{\rm dip}$ and $Ro$, for MS M-dwarfs if the sample is considered 
as a whole.  The latter trend is driven by the factor of $\sim$10 range of $P_{\rm rot}$ values across the sample rather than the range of 
$\tau_{\rm c}$ values, which vary by a factor of $\sim$2 (as listed in \citealt{don08a} and \citealt{mor08}), with the overall correlations
arising as all the observed fully convective stars (circles in Figure \ref{magrot} right panel) and almost all stars which are mostly convective 
(triangles) are faster rotators, while substantially radiative stars (asterisks) span a range of rotation rates.  

There is a clearer link between $B_{\rm dip}$ and $P_{\rm rot}$ for the PMS sample with the stars with the strongest 
dipole components (fully convective stars) spinning more slowly than stars with weaker dipole components 
(stars with radiative cores) - see Figure \ref{magrot}.  This is likely being driven by the star-disk interaction with stars with
stronger dipole components able to truncate their disks out to corotation, see \S\ref{dipcomp}, and those with weaker dipole
components having smaller disk truncation radii and therefore being spun-up.  GQ~Lup and MT~Ori (the rightmost three triangles 
in Figure \ref{magrot}) are not exceptions to this trend, as these stars have only recently ended the fully convective phase and 
presumably have not had enough time to spin-up.  This argument is supported by the strong correlation in Figure \ref{spinup} 
where we have estimated the time relative to the end of the fully convective phase [that is the age of a star minus 
the age at which a star of its mass is expected to develop a radiative core as calculated from equation 
(\ref{endfully})].  It is clear that the longer a star has spent with a radiative core, the faster its rotation period.\footnote{The trend in Figure 
\ref{spinup} is unlikely to be caused (at least entirely) by the stellar contraction, that is by the spin-up of stars as they contract with age in order to 
conserve angular momentum - there is no clear trend between $P_{\rm rot}$ and $R_{\ast}$ across our sample.  Furthermore, stars in 
Figure \ref{spinup} on either side of the fully convective divide span a range of ages.}  

We suggest that the relation between the stellar rotation rate and the strength of the dipole component for PMS stars is driven by the 
star-disk interaction rather than the dynamo magnetic field generation process itself.  This is further supported by the lack of any clear
relation between $B_{\rm dip}$ and Rossby number, see Figure \ref{magrot} (lower left panel).  Although $\tau_{\rm c}$ values drop from
a couple of hundred days to a few tens of days with the development of a large radiative core (e.g. \citealt{jun07}), $P_{\rm rot}$ 
values are also smaller for stars with large cores (see the asterisks in Figure \ref{magrot}, upper left panel).  Thus there is little variation in $Ro$
across our PMS sample, all of which lie in the saturated regime of the rotation-activity relation.  Likewise, there is little variation
in $Ro$ values for MS M-dwarfs that lie in the saturated regime ($Ro\le0.1$). 


\subsection{Non-accreting pre-main sequence stars}\label{nonaccn}
Given the importance magnetic fields in controlling the star-disk interaction, and in turn the stellar rotational evolution,
we have thus far focussed our discussion on accreting T Tauri stars.  Magnetic maps have also been published for one non-accreting 
weak line T Tauri star, V410~Tau \citep{ske10}; and a handful of post T Tauri stars which have long since lost their disks and have spun-up, 
HD~155555 (a close binary system; \citealt{dun08}), HD~141943 \citep{mar11}, and 
HD~106506 \citep{wai11}. With the exception of V410~Tau all of the non-accreting stars have substantial radiative cores
($M_{\rm core}/M_\ast\approx0.93$ and $\approx0.84$ for the primary and secondary stars of HD~155555) or 
have entirely radiative interiors (HD~106506 and HD~141943) as inferred from the models of \citet{sie00}.  
The non-accreting T Tauri stars are typically faster rotators 
than the accreting stars considered in \S\ref{topo} with $P_{\rm rot}\le2.2\,{\rm d}$, as is commonly found (e.g. \citealt{bou93}).  The 
post T Tauri stars have small, or no, outer convective zones and are found to host highly complex magnetic fields with many 
high order field components.  This is consistent with the magnetic evolutionary scenario discussed above for accreting 
T Tauri stars.  

V410~Tau is the only non-accreting star with a small radiative core for which magnetic maps have been 
obtained \citep{ske10}.  It is a young ($\sim$1.7$\,{\rm Myr}$) and higher mass star ($M_\ast\approx1.4\,{\rm M}_\odot$). 
Widely varying estimates of its effective temperature have been reported in the literature, and its luminosity is
highly uncertain given the large spot coverage (see the discussion in \citealt{ske10}).  Thus the position of V410~Tau
in the HR diagram is poorly constrained.  According to the models of \citet{sie00} this star has 
already ended the fully convective phase of evolution 
and has a small radiative core ($M_{\rm core}/M_\ast\approx0.07$), although given the uncertainty in its HR diagram position 
it may have an internal structure that ranges from fully convective, to having a moderately sized core 
($M_{\rm core}/M_\ast\approx0.3$; \citealt{ske10}).
Its complex magnetic field topology has more in common with the accreting T Tauri stars with large radiative cores, suggesting that 
it has indeed ended the fully convective phase.  However, with only one genuine weak line T Tauri star studied thus far 
it is not clear if the magnetic fields of these stars will follow similar trends as found for accreting classical T Tauri stars; but given 
that accreting stars (of age a few Myr) follow a similar magnetic topology trend 
with internal structure as found for MS M-dwarfs (of age a few Gyr), it is reasonable to assume that the magnetic topologies 
of more evolved PMS stars will follow suit.  If they do not, it may indicate that accretion is modifying the 
stellar magnetic field generation process.  Furthermore, if our conclusion that it is the star-disk interaction that is driving the 
relation between $P_{\rm rot}$ and $B_{\rm dip}$ (see \S\ref{tau_c} and Figure \ref{magrot}, upper left panel) is correct 
then we do not necessarily expect to find the same behavior for systems where the disk has dispersed.  With the influence 
of the disk removed, underlying relationships between the stellar magnetic field topology and the dynamo properties may be revealed.    
Non-accreting T Tauri stars will be the target of future spectropolarimetric observing campaigns to specifically address 
such issues.  


\section{Conclusions}\label{conclusions}
Spectropolarimetric observations carried out over at least a full stellar rotation, and ideally several rotation periods, combined with 
tomographic imaging techniques have allowed maps of the magnetic fields of a small sample of accreting T Tauri stars to be 
derived \citep{don07,don08b,don10a,don10b,don11a,don11b,don11c,don12,hus09,ske11}.  T Tauri
magnetic field topologies are found to vary with the stellar parameters.  We find that the large scale
topology appears to be directly linked to the internal structure of the star, and in particular to the size of the radiative
core that develops at the end of the fully convective phase of evolution.  

We define four regions across the HR diagram, see \S\ref{canwepredict} and Figure \ref{color_hr}, delineating stars with 
different magnetic topology characteristics.  Stars with substantial radiative cores, $M_{\rm core}/M_\ast\gtrsim0.4$, 
have complex fields that are highly non-axisymmetric with weak dipole components, only a few tenths of a kG at most
(this defines region 1 of the HR diagram as discussed in \S\ref{canwepredict} and colored blue in Figure \ref{color_hr}). 
V4046~Sgr~AB, CR~Cha and CV~Cha have this type of field topology.  Stars which have small radiative cores,
$0<M_{\rm core}/M_\ast\lesssim0.4$, have largely axisymmetric large scale magnetic field topologies but field modes
of higher order than the dipole component dominate.  Their dipole components are generally weaker than those found for fully convective
stars and appear to range from less than $0.1\,{\rm kG}$ to around $1\,{\rm kG}$.  
MT~Ori, TW~Hya, V2129~Oph, and GQ~Lup possess magnetic fields like this.
Intriguingly, stars that fall in this region of the HR diagram (region 2 discussed
in \S\ref{canwepredict} and colored green in Figure \ref{color_hr}) for which magnetic maps have been published 
all have strong octupole components to their magnetic fields, and this is the dominant field mode in TW~Hya, V2129~Oph, 
and GQ~Lup.  There are currently no theoretical models to explain this trend.     

We emphasize that the limit of $M_{\rm core}/M_\ast\approx0.4$ 
between the regions 1 and 2 is empirical and more observations are required to properly determine the exact boundary.  The boundary 
may also be a function of stellar mass itself, but we note that for main sequence M-dwarfs (whose topology trends with stellar
internal structure mirror the behavior of PMS stars) the transition from dominantly axisymmetric to dominantly non-axisymmetric
fields occurs at $M_\ast\sim$0.5$\,{\rm M}_\odot$ (see Figure \ref{mdwarfs}) which (roughly) corresponds to $M_{\rm core}/M_\ast\approx0.4$.

Fully convective stars of mass $\gtrsim0.5\,{\rm M}_\odot$ host simple fields that are dominantly axisymmetric
with strong dipole components of order one to a few kG (region 3 of the HR diagram as discussed
in \S\ref{canwepredict} and colored yellow in Figure \ref{color_hr}).  AA~Tau and BP~Tau fall into this category.  Such stars will develop
radiative cores before they arrive on the main sequence, at which point it seems likely that the dipole component of their magnetic fields 
will decay, but initially the axisymmetric nature of their fields will be maintained.  Further core growth will eventually destroy the 
axisymmetry of their magnetic fields leaving the stars with complex fields with weak dipole components.  It appears as though
this drop in the dipole component, which will reduce the disk truncation radius and increase the spin-up torque on the star, 
influences the stellar rotation rate with stars which have spent longer with radiative cores being faster rotators.   

Although the magnetic topology trends that we have observed across the PMS of the HR diagram (see 
Figures \ref{hr} \& \ref{color_hr}) are thus far based on a limited sample of stars, it is the overall excellent agreement between
the large-scale field topologies of PMS stars and those of MS M-dwarfs with comparable internal structures (i.e. similar ratios of core 
mass to stellar mass $M_{\rm core}/M_\ast$) that gives us
confidence to define distinct magnetic topology regimes.  Further spectropolarimetric studies of PMS stars across the HR diagram
are now required to test our conclusions.  By far the clearest trends, observed for both the MS and PMS sample, is the rapid increase
in field complexity, and the rapid decrease in the dipole component, when moving from objects close to the fully convective divide to those
with substantial radiative cores.

The lowest mass M-dwarfs (below $\sim$0.2$\,{\rm M}_\odot$, or later than spectral type $\sim$M5) are found to host a variety 
of field topologies which may be due to a bistable dynamo process \citep{mor10,mor11}.  
The similarity between the field topologies of MS M-dwarfs and PMS stars allows us to predict that bistable dynamo behavior, and therefore
stars with a variety of large scale field topologies, will be found for the lowest mass PMS stars.  We have thus defined
a fourth region of the HR diagram where such stars will be found.  The exact boundary between this region 4 and region 3 
is poorly constrained and more ZDI studies are required for stars in this low mass regime.  The only star studied in this region
thus far is V2247~Oph, a fully convective star with a complex field that likely resides on the weak field bistable dynamo branch
(stars on the strong field branch would host simple fields).  

Although he magnetic topology trends with stellar internal structure are empirical and currently lack theoretical grounding there
is additional evidence for a field topology change as stars transition from fully to partially convective.  With the growth of 
a radiative core large-scale magnetic fields become more and more complex.  \citet{sau09} argued that such a change
could explain their observed reduction in the number of periodically variable PMS stars, with fully convective stars hosting 
simple fields with large cool spots and partially convective stars more complex fields with more numerous and distributed smaller spots
(which causes less photometric rotational variability).   This is fully consistent with the Doppler maps of fully convective T Tauri 
stars which often show large (usually slightly offset from the rotation pole) high latitude spots (e.g. \citealt{don10b}).
Likewise \citet{reb06} and \citet{may10} find a systematic reduction
in the ratio of X-ray to bolometric luminosity, which is driven by the strength of the convective dynamo, for stars with
radiative cores.  Furthermore \citet{ale12} argue that the change from fully to partially convective 
stellar interiors can explain their observation that the scatter in X-ray luminosities in rotation-activity plots reduces with increasing  
PMS cluster age.  Whilst these observations lack the clarity or precision of our own they do add significant support to the change 
in magnetic field topology that we observe.  

We conclude that it is possible to predict the general characteristics of the magnetic field of a PMS
star based purely on its position in the HR diagram. For example whether the field will be axisymmetric with a strong 
dipole component, or axisymmetric with a field component of higher order than the dipole dominant, or complex and non-axisymmetric
with a weak dipole component.  Large scale magnetic field topologies are likely variable over time too, as has been observed for V2129~Oph, 
TW~Hya (tentatively), and GQ~Lup \citep{don11a,don11c,don12}.  However, although the polar strength of the various field components 
was observed to 
vary for all of these stars, the general characteristics of their large-scale fields remained the same at both epochs - dominantly
octupolar and well described by a tilted dipole plus a tilted octupole component \citep{gredon11}.  Likewise 
the general properties of the magnetic fields of AA~Tau and BP~Tau remained the same as derived from 
datasets taken in different observing seasons.    

We do caution, however, that our ability to predict the general large scale magnetic topology characteristics of a given PMS star 
is reliant on the veracity of the PMS evolution models themselves and on our ability to accurately position the star in the HR diagram
in the first place.  An alternative point of view, however, is that we can use ZDI studies and the derived magnetic topologies of T Tauri stars as a 
direct test of the internal structure information derived from the 
PMS evolutionary models themselves, just as dynamical mass measurements can constrain the mass tracks \citep{hil04} and 
lithium depletion the isochrones \citep{pall05}. For example, if a star is found
to host a largely axisymmetric field, but with dominant high order components, it is likely that this star has already ended
the fully convective phase of evolution and has developed a small radiative core.  If the star falls in a region of the 
HR diagram where fully convective stars lie, it could then be argued that either the models are inadequate in this region,
and/or a better assignment of the stellar effective temperature and luminosity needs to be made.  

In principle ZDI studies can be used to provide strong observational constraints on the divide between the fully and 
partially convective regions of the PMS in the HR diagram; a divide that is exquisitely model dependent. 
Pinpointing this divide observationally should allow the detailed testing of evolutionary models of stellar internal
structure.  Observationally constraining the fully convective divide as a function
of mass and age does not only have important implications for periodic variability and X-ray emission, as discussed above.
Additionally the development of a radiative core likely leads to a dramatic redistribution of angular momentum as the 
convective envelope and core decouple \citep{end81}, which, if further studied, should yield insights into such phenomena as 
rotationally induced mixing (e.g. \citealt{pin97}).   Furthermore, as the interaction with a circumstellar disk is (in most, but crucially 
not all, cases) dominated by the large scale dipole component of the magnetic field \citep{gre08,ada11}, 
studying stars with disks across the fully convective/radiative core divide will enable us to probe the dynamics and physics of 
magnetopsheric accretion as a function of magnetic field topology in an extremely targeted fashion.

Data from the Magnetic Protostars and Planets (MaPP) program will continue to be obtained until at least 
the end of 2012.  This continued stream of T Tauri magnetic maps, coupled with
those for stars already observed as part of MaPP but not yet published, will allow the HR diagram to be more 
fully populated.  This will allow the boundaries separating the different topology regions within the HR diagram 
to be better constrained observationally.  Furthermore, by repeatedly observing the same stars over timescales
of several years we will gain insight into the long term variability of the large scale magnetospheres of T Tauri stars,
and possibly the existence of magnetic cycles.  MaPP will thus further advance our understanding of the 
magnetism of forming low mass, including solar-like, stars.  From a theoretical perspective models of the 
magnetospheric accretion process that incorporate magnetic fields with an observed degree of complexity 
have been developed \citep{gre05,gre06a,gre08,gre10,lon08,lon11,lon12,moh08,rom11,gredon11,ada11}.  However, the observed
variations in the magnetic field topology with the development of a radiative core, and the possible bistable
dynamo process that appears to operate amongst the lowest mass T Tauri stars, highlight the need
for new models, similar to those of \citet{mat10,mat12}, but which take proper account of both the magnetic field complexity, 
and the field variation with changes in the stellar internal structure.  


\acknowledgments
The authors thank L. Siess, N. Baliber \& F. C. Adams for insightful discussions, E. Tognelli \& P. G. Prada Moroni for sending 
stellar internal structure information from their PMS evolution models, Y.-C. Kim for sending convective 
turnover time estimates, and the referee for their useful comments. SGG is supported by NASA grant HST-GO-11616.07-A. JM is supported by a postdoctoral fellowship 
of the Alexander von Humboldt foundation. The ``Magnetic Protostars and Planets'' (MaPP) project is supported by the 
funding agencies of CFHT and TBL (through the allocation of telescope time) and by CNRS/INSU in particular, as well as by the French
``Agence Nationale pour la Recherche'' (ANR). 


\appendix

\section{Accreting T Tauri stars with magnetic maps derived from ZDI}\label{starinfo}
\subsection{Stars with strong dipole components and axisymmetric large scale magnetic fields}
\subsubsection{AA Tau}
AA~Tau is one of the best studied accreting stars and hosts the simplest large scale magnetic field yet discovered on
any T Tauri star.  Its large scale field is dominantly dipolar, with weak high order field components \citep{don10b}.  The dipole component 
is strong with a polar field strength of $\sim$2~kG (perhaps a large as 3~kG, see the discussion in \citealt{don10b}) 
and is tilted by $\sim$10--20$^{\circ}$ with respect to the stellar rotation
axis, see Table \ref{params}.  The star was observed at two different epochs separated by a year. The large scale field topology showed 
no significant evolution, which may be linked to the moderate phase coverage obtained at both epochs, but repeat observations 
are required to confirm this.  Given the weak mass accretion rate on to AA~Tau during the ESPaDOnS observations
(an average of $\log\dot{M}=-9.2\,{\rm M}_\odot{\rm yr}^{-1}$; \citealt{don10b}) coupled with the strength of the dipole 
component the disk may be truncated close-to the equatorial coorotation radius, or even beyond at some 
epochs.  The mass accretion rate, however, was observed to vary by an order of magnitude 
from $\log\dot{M}=-9.6$ to $-8.5\,{\rm M}_\odot{\rm yr}^{-1}$.  AA~Tau has a completely convective interior according to 
both the \citet{sie00} and the \citet{tog11} PMS stellar evolution models.


\subsubsection{BP Tau}\label{bptau}
BP~Tau has long been known to possess a strong stellar-disk averaged magnetic field from Zeeman broadening 
measurements \citep{joh99a}, with strong circular polarization measured in the accretion related HeI 5876{\AA } 
emission line \citep{joh99b,sym05,chu07}.  Like AA~Tau, ZDI has revealed that BP~Tau has a strong dipole component
to its magnetic field \citep{don08b}.  However it also possesses a strong 
octupole field component (of polar field strength of 1.6-1.8~kG compared to the dipole component of polar field strength 
1-1.2~kG; see footnote c of Table \ref{params}).  
Both the dipole and octupole moments are tilted relative to the stellar rotation axis, but by different amounts and towards different 
rotation phases.  Magnetic maps have been derived at two different epochs, separated by around 10 months \citep{don08b}.  The large
scale field topology showed little change over this time, apart from an apparent rotation of the entire surface field by 0.25 in phase.  
This was most likely caused be a small error in the assumed rotation period (7.6$\pm$0.1~d) building up over the $\sim$39 rotations 
between the observing epochs.  Variations in the large scale field topology cannot, however, be ruled out on longer time scales.  
We further note that the magnetic maps for BP~Tau were published 
prior to the MaPP project, using an experimental version of the magnetic imaging code \citep{don08b}.  This code considered 
polarization signals in the photospheric absorption lines and the accretion related emission lines separately, whereas the more mature
version of the code constructs the maps by considering both signals simultaneously \citep{don10b}.  The archival spectropolarimetric
observations of BP~Tau will be re-analyzed in a forthcoming paper, and will be presented alongside new recently obtained data. 
From its position in the HR diagram BP~Tau is a fully convective star.
   

\subsection{Stars with dominant high order magnetic field components and axisymmetric large scale fields}
\subsubsection{V2129 Oph}
V2129~Oph was the first accreting T Tauri star for which magnetic maps were published (\citealt{don07}; see also \citealt{don11a} 
for a re-analysis of the original dataset using the latest version of the magnetic imaging code).  It has been observed at two
different epochs, June 2005 and July 2009 \citep{don07,don11a}.  At both epochs V2129~Oph was found to
host a dominantly octupolar magnetic field. The dipole component of its multipolar magnetic field was found to vary
by a factor of about three, from $\sim$0.3$\,{\rm kG}$ to $\sim$0.9$\,{\rm kG}$, in the four years between the observing runs \citep{don11a}.  The clear
detection of secular evolution of the large scale magnetic field demonstrates that it is dynamo generated and not of fossil 
origin.  At both epochs the dipole and octupole field components were found to be slightly tilted with respect to the stellar rotation axis 
and tilted towards different rotation phases.  V2129~Oph has a binary companion \citep{ghe93}
although this is about 50 times fainter in the V-band than V2129~Oph itself \citep{don07}.  The projected separation of
0.65'', as measured by \citep{cie10}, translates to $78\,{\rm AU}$ assuming a distance of $120\,{\rm pc}$ to the $\rho$ Oph star forming 
region \citep{loi08}.  V2129~Oph is no longer fully convective and has developed a small radiative core, $M_{\rm core}/M_\ast\approx0.2$ \citep{sie00}. 


\subsubsection{GQ Lup}\label{gqlup}
GQ~Lup was found to host a dominantly octupolar magnetic field when observed in both July 2009 and June 2011 \citep{don12}.  However
its large scale field weakened considerably between the two observing epochs indicating a non-stationary dynamo
process.  The polar strength of the octupole component dropped from $2.4\,{\rm kG}$ to $1.6\,{\rm kG}$ and that of 
the dipole from $1.1\,{\rm kG}$ to $0.9\,{\rm kG}$ between 2009 and 2011.  At both epochs the octupole was 
roughly aligned with the stellar rotation axis but the dipole component was tilted by $\sim30^{\circ}$ \citep{don12}.  
Of the stars for which magnetic maps have been derived to date,
GQ~Lup displays the strongest large scale fields yet discovered, with the longitudinal field component measured in accretion
related emission lines (i.e. those which probe the field where accretion columns impact the star) reaching $6\,{\rm kG}$.  GQ Lup has 
a known substellar companion in its outer accretion disk, orbiting at $\sim100\,{\rm AU}$ \citep{neu05}, that is 
most likely a brown dwarf \citep{lav09}.  GQ~Lup has developed a small radiative core, $M_{\rm core}/M_\ast \approx 0.13$ \citep{sie00}.


\subsubsection{TW Hya}\label{twhya}
As with V2129~Oph and GQ~Lup, TW~Hya was found to host a dominantly octupolar magnetic field at the two epochs it
was observed \citep{don11c}.  At both epochs the dipole component was found to be weak relative to the octupole component, with
the ratio of their polar strengths varying from $B_{\rm oct}/B_{\rm dip}\approx6$ in March 2008 to $B_{\rm oct}/B_{\rm dip}\approx4$ 
in March 2010.  At the first epoch the positive pole of the dipole component was found to be tilted by about 45$^{\circ}$ 
relative to the main negative pole of the octupole component.
At the second epoch the dipole and octupole moments were roughly anti-parallel, with the main negative pole of the octupole coincident
with the visible rotation pole of the star.  The change in the tilt of the dipole component is tentative, however, given the relative weakness of the 
dipole component and the limited phase coverage obtained during the first observing run \citep{don11c}.  It may be indicative of a magnetic
cycle, but clearly repeated observations, potentially over many years and with improved phase coverage, are required to confirm this. 
TW~Hya is a somewhat atypical T Tauri star 
given its low inclination ($\approx$7$^{\circ}$; \citealt{qi04}) and since it still has a significant mass accretion rate 
(averaging $\log\dot{M}=-8.9\,{\rm M}_\odot{\rm yr}^{-1}$ at both observing epochs; \citealt{don11c}) despite being of an age ($\sim9\,{\rm Myr}$)
where the disks of most T Tauri stars have dispersed and accretion has ceased (e.g. \citealt{fed10}).  TW~Hya 
has an interior structure that consists of a radiative core surrounded by an outer convective envelope, with a core mass of 
$M_{\rm core}/M_\ast\approx0.2$ \citep{sie00}.


\subsubsection{MT Ori}
MT~Ori hosts a complex magnetic field with the surface of the star covered in many 
regions of opposite polarity, although its large scale field is dominantly axisymmetric (i.e. the $m=0$ field modes dominate; \citealt{ske11}).  
The large scale dipole component was found to be weak $<100\,{\rm G}$ with the field dominated by the octupole ($\ell=3$), the dotriacontapole
($\ell=5)$, and the $\ell=7$ field modes.  The total contribution from the $3\le\ell\le7$ field components was 13 times stronger
than the dipole ($\ell=1$) component with the octupole four times stronger than the dipole \citep{ske11}.
At $\sim$2.7$\,{\rm M}_\odot$ according to the models of \citet{sie00}, or $\sim$2$\,{\rm M}_\odot$ using the models of \citet{tog11}, and 
$\sim$0.25$\,{\rm Myr}$ this is the highest mass and youngest star in the MaPP sample.    
Despite its young age, MT~Ori is massive enough to have already developed a small radiative core, at least in the 
models of \citet{sie00}.  The \citet{tog11} models suggest that MT~Ori is still fully convective.  Given the similarity of its magnetic field to 
that of TW~Hya, V2129~Oph, and GQ~Lup, stars which have small radiative cores in both PMS evolution models, and the dissimilarity 
between its field and the 
simple fields of the fully convective stars AA~Tau and BP~Tau, we suggest that MT~Ori does indeed have a small radiative core.  
Given the large uncertainty in its effective temperature and luminosity (see \citealt{ske11}) the core mass lies 
somewhere in the range $0.01\le M_{\rm core}/M_\ast\le0.36$ according to the 
\citet{sie00} models.         


\subsection{Stars with complex non-axisymmetric large scale magnetic fields with weak dipole components}

\subsubsection{V4046 Sgr AB}
Both stars of the close binary system V4046~Sgr host complex magnetic fields with many high order field components \citep{don11b}.  
The large scale magnetosphere of each star, their dipole components, are weak and highly tilted with respect to their rotation 
axes (of polar strength $\sim$100$\,{\rm kG}$ and 
$\sim$80$\,{\rm kG}$ and tilted by 60$^{\circ}$ and 90$^{\circ}$ on the primary and secondary respectively).  
The planes of the tilts 
of the dipole moments are also offset by roughy 0.7 in rotation phase, further increasing the field complexity.  The binary 
magnetospheric structure is highly complex and will be presented in a future paper.  A circular polarization signal was not detected
in the accretion related emission lines, consistent with the complex magnetic geometries and likely indicative of accretion 
spots being distributed across many opposite polarity regions \citep{don11b}.    
The binary orbit is circularized and
synchronized with accretion occurring from a circumbinary disk \citep{ste04,rod10}.  Recent numerical simulations suggest that small
local circumstellar disks, distinct from the global circumbinary disk, may also form around the individual stars \citep{dev11}.  
As with TW~Hya (see \S\ref{twhya}), V4046~Sgr (age $\sim$13 Myr) is still accreting at an age when most T Tauri stars 
have lost their disks (e.g. \citealt{fed10}).   
The masses of V4046~Sgr~AB 
listed in Table \ref{params} are derived from the PMS evolution models and placing the stars on the HR diagram.  These can be compared to 
the more accurate dynamical masses of $0.912\,{\rm M}_\odot$ and $0.873\,{\rm M}_\odot$ calculated by \citet{ste04} for 
V4046~Sgr~A and V4046~Sgr~B respectively.  Both binary components have ended the fully convective phase of evolution with
the primary and secondary have core masses of roughly 50$\%$ and 40$\%$ of their respective stellar masses using the models of \citet{sie00}.  


\subsubsection{CR Cha}
CR~Cha hosts a particularly complex magnetic field with a significant fraction of the magnetic energy in high $\ell$-number field
modes \citep{hus09}.  Unlike the other T Tauri stars discussed in this paper CR~Cha (and CV~Cha, see below) were observed
with SemelPol, a spectropolarimeter at the Anglo-Australian telescope.  As with V4046~Sgr, a circular polarization 
signal was not detected in the accretion related emission lines.  \citet{hus09} noted that this non-detection may have been due to 
insufficient S/N.  CR~Cha is too far south to be re-observed with ESPaDOnS.  However, as Stokes V signals were not detected 
in the emission lines in the higher S/N ESPaDOnS spectra of V4046~Sgr, and as the magnetic maps of V4046~Sgr~AB and CR~Cha
reveal a similar level of field complexity, a more likely explanation is that magnetospheric accretion on to the surface 
of CR~Cha occurs into several distributed opposite polarity magnetic regions, yielding a net polarization signal of zero due to the 
flux cancellation effect.  Because of this the value listed for the dipole component of the multipolar field of CR~Cha in Table \ref{params} 
is a lower limit and the true value may be larger.   
According to the \citet{sie00} models it has a substantial radiative core 
$M_{\rm core}/M_\ast = 0.65$.


\subsubsection{CV Cha}
The magnetic field of CV~Cha is highly complex with many high order field components \citep{hus09}.   As with CR~Cha a 
circular polarization signal was not detected in the accretion related emission lines, and the dipole component listed in 
Table \ref{params} is likely a lower limit to the true value.  CV~Cha is the primary star of a large separation 
(1596$\,{\rm AU}$) binary system; CW~Cha being the secondary star \citep{rei93}.   
Of the two Chamealeon I stars for which magnetic maps have been published, CV~Cha has the larger mass accretion rate, 
$\log{\dot{M}}=-7.5\,{\rm M}_\odot {\rm yr}^{-1}$ compared to $\log{\dot{M}}=-9.0\,{\rm M}_\odot {\rm yr}^{-1}$ for 
CR~Cha \citep{hus09}. 
CV~Cha is the earliest spectral type star in our sample (see Table \ref{params_fund}), 
is well into the Henyey phase of its evolution, and is almost entirely radiative ($M_{\rm core}/M_\ast \approx 1.0$; \citealt{sie00}). 
 

\subsubsection{V2247 Oph}\label{v2247} 
V2247~Oph is the lowest mass T Tauri star for which magnetic maps have been published.  It is fully convective and it is found to host a 
complex magnetic field with a weak dipole component \citep{don10a}.  Its topology is therefore similar to the more massive T Tauri stars 
which have developed substantial radiative cores, rather than the simple fields of the other fully convective stars AA~Tau and BP~Tau.  
Its mass of $\sim$0.36$\,{\rm M}_\odot$ has been obtained from the \citet{sie00} models using a luminosity appropriate for a distance
to the $\rho$ Oph star forming region of $120\,{\rm pc}$ \citep{loi08}.  V2247~Oph is weakly accreting and has 
previously been classified as a non-accreting weak line T Tauri star (e.g. \citealt{bou92}).  However, all of the accretion related 
emission lines are present in the ESPaDOnS spectra, with evidence for a weak accretion rate.  The accretion rate is found to 
be highly variable over timescales of order a week \citep{don10a}, and over several years \citep{lit04}, reaching peaks of 
around $\log\dot{M}\approx{-9}\,{\rm M}_\odot{\rm yr}^{-1}$.  The spectral energy distribution of V2247~Oph suggests that its disk is rather 
evolved with a large inner (dust) disk gap \citep{gra05}.  This may be due to a nearby binary companion star \citep{sim87}, 
separated by $36\,{\rm AU}$ adopting the distance given above, suggesting that accretion occurs from a circumbinary 
disk.  V2247~Oph is rotating about twice as fast ($P_{\rm rot}\sim3.5\,{\rm d}$) as the other
fully convective stars in the sample, AA~Tau ($P_{\rm rot}=8.22\,{\rm d}$) and BP~Tau ($P_{\rm rot}=7.6\,{\rm d}$).  As speculated
by \citet{don10a} the faster spin rate of V2247~Oph may be a direct reflection of its complex magnetic field.  Stars with magnetic 
fields with weaker dipole components would have disks that are magnetospherically truncated closer to the star, potentially
resulting in a larger spin-up torque in comparison to that experienced by stars with stronger dipole components that are able to truncate
their disks at larger radii (e.g. \citealt{mat05}).  The variation in field topology between the more massive fully convective PMS stars 
and the low mass PMS star V2247~Oph is similar to what has been found for the main sequence M-dwarfs, see \S\ref{lowmass} and
\citet{mor10,mor11}.


\section{The fully convective limit as a function of stellar age}\label{fullylimit}

\begin{figure}[t]
  \centering
    \includegraphics[width=65mm]{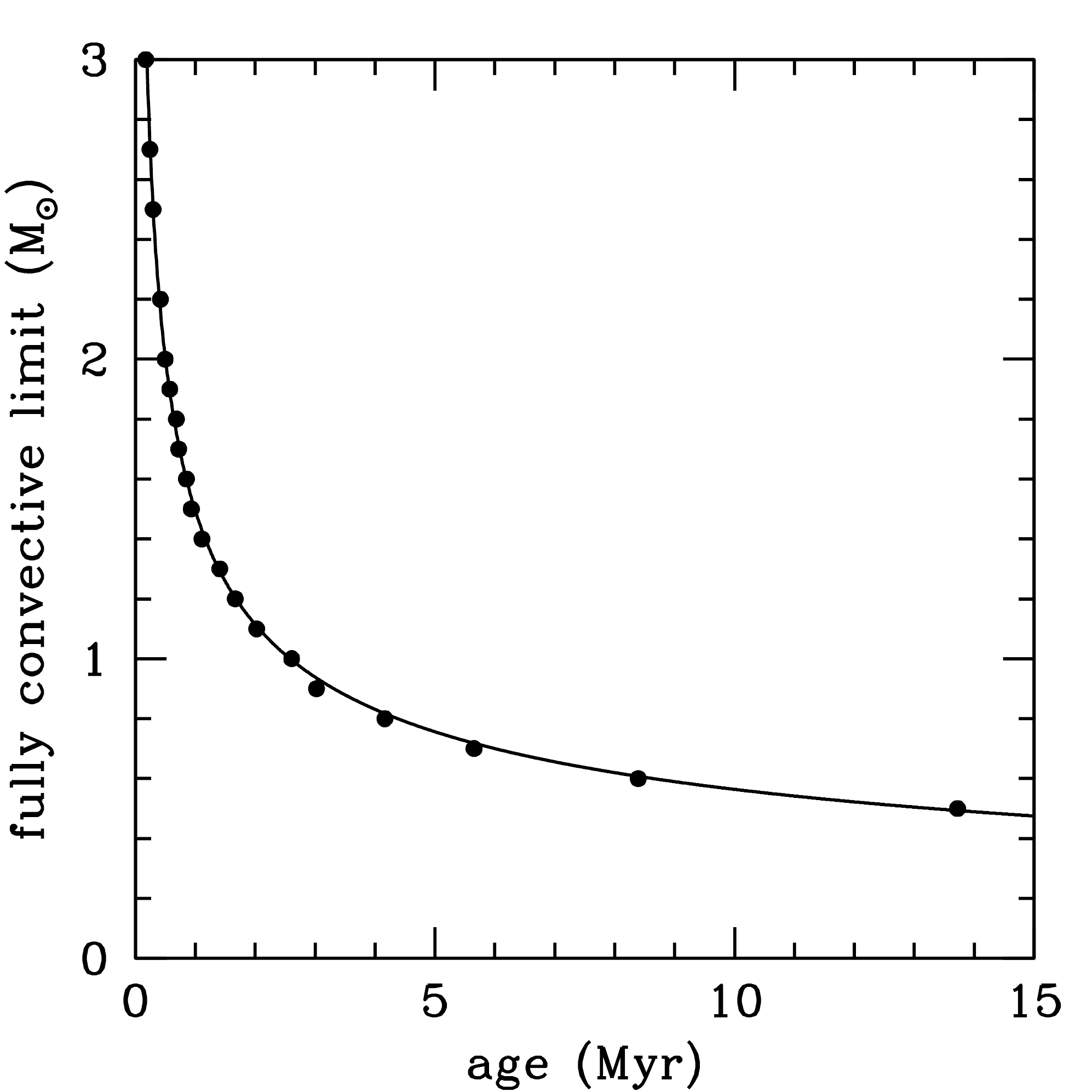}
        \caption{The fully convective limit as a function of age.  A power-law fit (solid line) to the
               data from the \citet{sie00} models (points) is given by equation (\ref{endfully}).  
               At a given age stars of higher mass than the fully convective limit have developed radiative cores
               while those of lower mass remain fully convective.}
               \vspace{10mm}
  \label{age_core2}
\end{figure}

The age at which a star ends the fully convective phase of evolution (the fully convective limit) is a function of
stellar mass (see Figure \ref{age_core2}).  From the \citet{sie00} models ($Z=0.02$ with convective overshooting) 
we can estimate the age at which the fully convective phase ends,   
\begin{equation}
{\rm age}\,[{\rm Myr}] \approx \left( \frac{1.494}{M_\ast/{\rm M}_\odot}\right )^{2.364}, 
\label{endfully}
\end{equation}
which is the power-law fit, the solid black line, in Figure \ref{age_core2} (stars below $\sim$0.35$\,{\rm M}_\odot$
remain fully convective e.g. \citealt{cha97}).  Thus a $0.5\,{\rm M}_\odot$ star ends 
the fully convective phase and develops a radiative core at an age of $\sim$13.3$\,{\rm Myr}$, while a    
$1\,{\rm M}_\odot$ and a $2\,{\rm M}_\odot$ star develop a radiative core at an age of 
$\sim$2.6$\,{\rm Myr}$ and $\sim$0.5$\,{\rm Myr}$ respectively.  The early development of a radiative core, and its more rapid
growth, for more massive stars leads to a gap in the observed color-magnitude diagrams of young PMS 
clusters. The size of which, as discussed by \citet{may07}, is a function of age and can possibly be used 
as a distance independent age indicator.   

\begin{figure}[t]
  \centering
    \includegraphics[width=65mm]{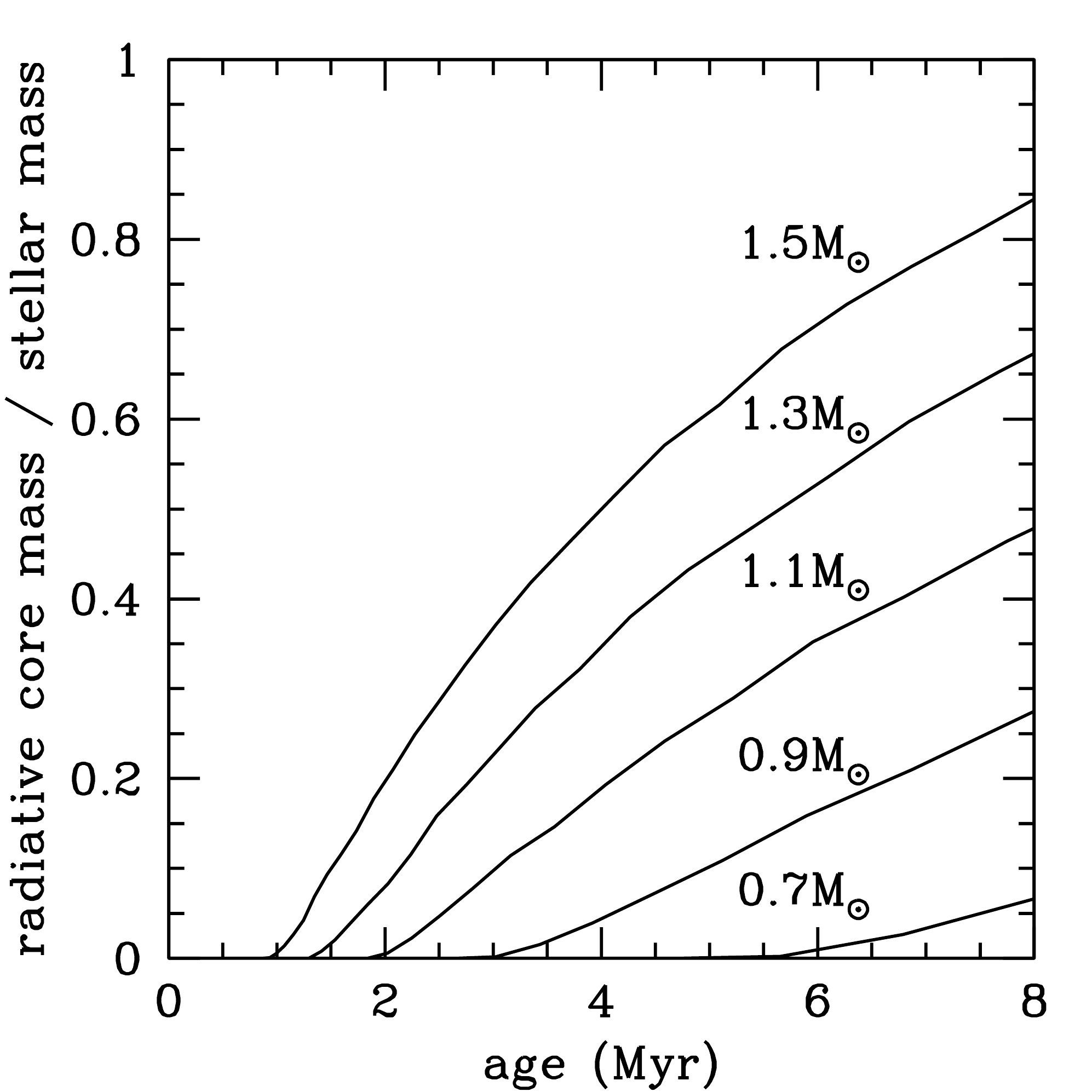}
    \includegraphics[width=65mm]{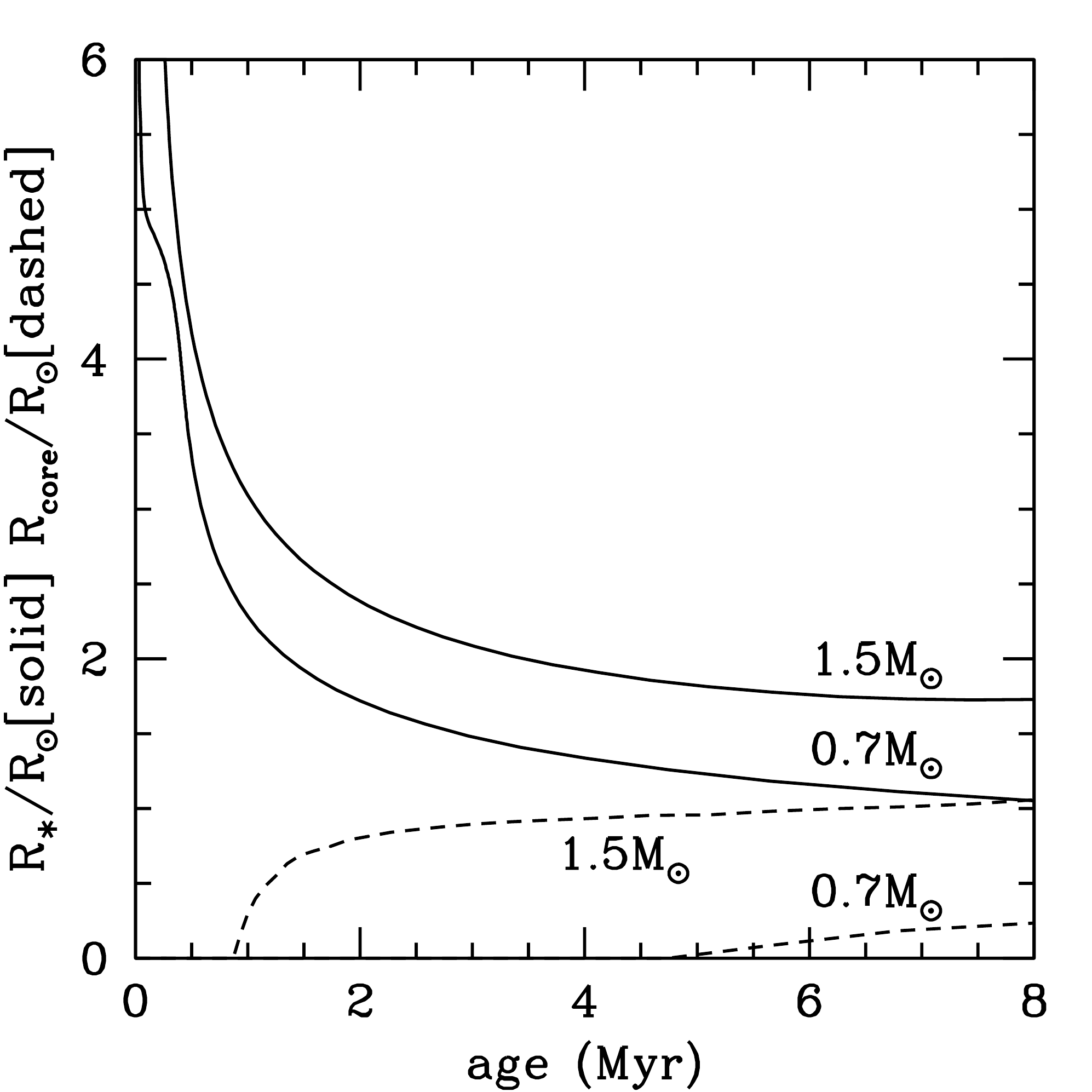}
  \caption{(left) The growth of the radiative core mass relative to the stellar mass as a function of age obtained from the 
               \citet{sie00} PMS stellar evolution models for stars of different mass.  An ordinate value of zero (one) represents a 
               star which has a fully convective (completely radiative) interior.  Higher mass T Tauri stars
               develop radiative cores at a younger age, 
               and the mass of the core relative to the stellar mass increases faster, compared to lower mass T Tauri stars. (right) The decrease in the
               stellar radius ($R_\ast$; solid lines) and increase in the core radius ($R_{\rm core}$; dashed lines) with age.  
               Only stellar radii below $6\,{\rm R}_\odot$, and 
               the behavior for a $1.5\,{\rm M}_\odot$ star and a $0.7\,{\rm M}_\odot$ star, are shown for clarity.  The brief slowing of the 
               contraction for the $0.7\,{\rm M}_\odot$ star at $\sim$0.5$\,{\rm Myr}$ is caused by deuterium burning; this occurs at an earlier 
               age and off the vertical scale for the $1.5\,{\rm M}_\odot$ star.                  
               }
  \label{age_core}
\end{figure}

Recently new mass tracks and isochrones have been published by \citet{tog11}; the Pisa PMS 
stellar evolution models.  The Pisa model equivalent of equation (\ref{endfully}) is     
\begin{equation}
{\rm age}\,[{\rm Myr}] \approx \left( \frac{1.448}{M_\ast/{\rm M}_\odot}\right )^{2.101}
\label{endfullypisa}
\end{equation} 
where $Z=0.02, Y=0.288, \alpha=1.68$, and $X_{\rm D}=2\times10^{-5}$ have been assumed.  Thus in the Pisa models
a $0.5\,{\rm M}_\odot$ star ends the fully convective phase and develops a radiative core at an age of 
$\sim$9.3$\,{\rm Myr}$, while a $1\,{\rm M}_\odot$ and a $2\,{\rm M}_\odot$ star develop a radiative core at an age 
of $\sim$2.2$\,{\rm Myr}$ and $\sim$0.5$\,{\rm Myr}$ respectively.  A comparison between equations (\ref{endfully}) and 
(\ref{endfullypisa}) reveals that the fully convective phase ends at approximately the same age for high
mass T Tauri stars, with differences of $<$0.5$\,{\rm Myr}$ for stars of mass $>$0.95$\,{\rm M}_\odot$, but this difference is 
larger for lower mass stars and exceeds $4\,{\rm Myr}$ for stars of mass $<$0.5$\,{\rm M}_\odot$.  
As lower mass stars are more likely to lose their
disk before a radiative core develops, and therefore before any drastic change in the large-scale field topology
occurs, then the difference in the age at which a core develops between the \citet{sie00} and the \citet{tog11}
models is not too significant.  For example, for a $0.5\,{\rm M}_\odot$ star the fully convective phase ends at $13.3\,{\rm Myr}$
or $9.3\,{\rm Myr}$ in the \citet{sie00} and \citet{tog11} models respectively.  However, the fraction of stars with disks (substantial, primordial,
dusty disks) is only $\sim$0.5\% or $\sim$2.4\% at such ages, using the disk lifetime estimate given by \citet{mam09}.  
Thus the $4\,{\rm Myr}$ year difference in age between the end of the fully convective phase in the two PMS evolution models
has little effect in terms of any stellar spin-up that may occur when a star is still coupled to its disk when a radiative core develops.
Of course, we must also account for the fact that for a given star the different mass tracks and isochrones of each 
model yield different age and mass estimates.


\end{document}